\documentclass[twocolumn]{article}

\usepackage[T1]{fontenc}
\usepackage[utf8]{inputenc}
\usepackage{lmodern}

\usepackage{amsmath, amssymb}
\usepackage{siunitx}
\usepackage[version=4]{mhchem}

\usepackage[gen]{eurosym}
\DeclareSIUnit{\EUR}{\euro}
\usepackage{adjustbox}
\usepackage{tabularx}
\usepackage{graphicx}
\usepackage{booktabs}
\usepackage{multirow}
\usepackage{caption}
\usepackage{float}
\usepackage{color, colortbl}
\definecolor{lightgray}{gray}{0.9}

\usepackage{geometry}
\usepackage[colorlinks=true, linkcolor=blue, citecolor=blue, urlcolor=blue]{hyperref}
\usepackage{url}

\usepackage{nomencl}
\makenomenclature
\usepackage[numbers]{natbib}
\title{\bfseries Enhancing Industrial Flexibility and Market Participation in Cement Manufacturing Through Optimized Production Scheduling}

\author{
\begin{tabular}{c}
Sebastián Rojas-Innocenti\textsuperscript{1}, Enrique Baeyens\textsuperscript{2}, Alejandro Martín-Crespo\textsuperscript{1},\\
Sergio Saludes-Rodil\textsuperscript{1}, Fernando Frechoso\textsuperscript{3} \\
\textsuperscript{1}CARTIF Technology Center, Valladolid, Spain \\
\textsuperscript{2}Universidad de Valladolid, Instituto de las Tec. Avanzadas de la Producción, Spain \\
\textsuperscript{3}Universidad de Valladolid, Departamento de Ingeniería Eléctrica, Spain \\
\texttt{sebroj@cartif.com}
\end{tabular}
}

\date{}

\begin{document}

\maketitle

\begin{center}
\begin{minipage}{0.95\linewidth}
\small
\textbf{Abstract.}
The growing share of variable renewable energy (VRE) sources in power systems is increasing the need for short-term operational flexibility—particularly from large industrial electricity consumers. This study proposes a practical, two-stage optimization framework to unlock this flexibility in cement manufacturing and support participation in electricity balancing markets. In Stage 1, a mixed-integer linear programming (MILP) model minimizes electricity procurement costs by optimally scheduling the raw milling subsystem. In Stage 2, a flexibility assessment model evaluates profitable deviations, targeting participation in Spain’s manual Frequency Restoration Reserve (mFRR) market. A real-world case study in a Spanish cement plant—including PV and battery storage—shows that flexibility services can yield monthly revenues of up to \EUR{}800 and paybacks as short as six years. This framework offers a replicable pathway for industrial flexibility in energy-intensive sectors.

\vspace{0.5em}
\textbf{Keywords:} Cement industry, demand response, operational flexibility, electricity balancing markets, MILP, battery storage, renewable integration.
\end{minipage}
\end{center}

\vspace{1em}

\section{Introduction}
\label{sec:intro}

The increasing penetration of variable renewable energy (VRE) sources—such as wind and solar—has intensified the need for operational flexibility to ensure power system stability. To balance real-time fluctuations in supply and demand, grid operators are increasingly turning to demand-side resources. Among these, industrial electricity consumers represent a promising, yet underutilized, source of flexibility. Energy-intensive sectors with partially controllable processes—such as cement manufacturing—are particularly well-positioned to contribute to this effort while pursuing energy cost reductions and decarbonization goals.

The cement industry alone accounts for 7--8\% of global anthropogenic \ce{CO2} emissions and nearly 7\% of global industrial energy use~\cite{IEA_cement,RTE_cement,BBC_cement,DW_cement}. Within the production chain, subsystems such as raw milling exhibit intrinsic flexibility due to their decoupling from the continuous kiln operation via intermediate silos. However, leveraging this potential in practice requires advanced decision-support tools capable of reconciling operational constraints with the volatility of electricity markets.

While demand-side response (DSR) programs offer a framework for incentivizing flexible industrial consumption, their implementation in continuous-flow environments remains challenging. Production schedules must satisfy strict requirements, including minimum machine uptime/downtime, inventory management, and uninterrupted material flow. Optimization-based methods are essential for identifying technically feasible and economically attractive scheduling adjustments.

This paper proposes a two-stage mixed-integer linear programming (MILP) framework to support market-responsive scheduling in cement manufacturing. The first stage determines a cost-optimal production schedule for the raw milling subsystem using day-ahead electricity prices. The second stage evaluates deviations from this baseline to provide upward and downward reserve capacity in Spain’s tertiary regulation market (manual Frequency Restoration Reserve, mFRR).

The methodology is validated through a real-world case study of a Spanish Portland cement plant. Using actual operational data and electricity market prices from 2023, the framework analyzes 19 asset configurations that combine photovoltaic (PV) generation and battery energy storage systems (BESS). Results quantify potential electricity cost savings, flexibility revenues, and investment payback times.

\subsection*{Contributions}

This study provides the following contributions:

\begin{enumerate}
    \item A two-stage MILP framework integrating cost-optimal production scheduling with short-term flexibility valuation under real market signals.
    \item A techno-economic assessment of 19 PV and BESS configurations, quantifying cost savings, flexibility revenues, and simple payback periods.
    \item Operational guidance for industrial stakeholders on monetizing demand-side flexibility with minimal disruption to production processes.
\end{enumerate}

\subsection*{Paper Organization}

The remainder of the paper is structured as follows. Section~\ref{sec:literature} reviews related work on industrial demand-side flexibility, with a focus on the cement sector. Section~\ref{sec:prob_stat} introduces the case study context. Section~\ref{sec:methodology} presents the two-stage optimization framework. Section~\ref{Case_study_main_chap} describes the simulation setup. Section~\ref{sec:results} discusses the results, and Section~\ref{sec:conclusion} concludes the paper and outlines future research directions.

\section{Literature Review}
\label{sec:literature}

This section reviews prior research on industrial demand-side flexibility, emphasizing applications in cement manufacturing. It highlights the role of optimization models and energy storage systems in enabling flexible operations and identifies key gaps addressed by this study.

\subsection*{Industrial Demand-Side Response}

Industrial demand-side response (DSR) is increasingly recognized as a critical enabler of power system flexibility~\cite{rollert_demand_2022, pierri_integrated_2020}. Numerous optimization-based approaches—including heuristics, model predictive control, and mathematical programming—have been developed to align industrial consumption with market signals while ensuring operational feasibility~\cite{boldrini_demand_2023, Zhao_2014}.

Recent studies emphasize the importance of modeling real-world constraints and enabling participation across multiple market layers. Rollert et al.~\cite{rollert_demand_2022} provide a comprehensive review of European industrial sectors, identifying heavy loads as the dominant source of technical flexibility. Boldrini et al.~\cite{boldrini_demand_2023} propose a multi-market MILP model for steelmaking, achieving significant cost reductions and balancing revenues. Adiguzel et al.~\cite{adiguzel_global_2024} conduct a global meta-analysis showing that industrial sites equipped with batteries outperform residential ones in up-regulation success rates. Rahmani et al.~\cite{rahmani_residential_2023} show that adding a 2~MWh BESS to a food processing facility increases accepted mFRR bids by 35\% without impacting performance metrics.

Together, these contributions underscore the need for application-specific, constraint-aware optimization frameworks—precisely the gap this work aims to address in the cement industry.

\subsection*{Grid-Responsive Operation in the Cement Industry}

Cement manufacturing presents significant short-term flexibility potential due to its high energy use and modular processes. In particular, raw milling and grinding stages are decoupled from the kiln through intermediate silos, enabling load shifting and schedule rescheduling~\cite{olsen_opportunities_2011, lee_evaluation_2020, rombouts_flexible_2021}.

Building on this foundation, Rombouts et al.~\cite{rombouts_flexible_2023} demonstrate that 24\% of electricity use is shiftable in a Belgian cement plant, though flexibility is not monetized. Parejo-Guzmán et al.~\cite{parejo_guzman_methodological_2022} reduce energy costs in Spanish clinker production by 11\% using a genetic algorithm under time-of-use tariffs. Zhang and Wu~\cite{zhang_mpc_2023} develop a rolling MILP with model predictive control to enable flexibility participation in China's AGC market. Tong et al.~\cite{tong_pv_bess_2024} show that coupling PV with BESS increases profitable mFRR windows by 60\% at a North African plant.

Despite these advances, few studies jointly optimize production schedules and flexibility participation under real-time electricity market signals. Most omit a comparison of flexibility revenues against baseline operational costs and overlook the synergy between production planning and storage assets. Additionally, reserve market participation remains underexplored in the cement context.

\subsection*{Research Gaps and Originality of This Study}

This study addresses the aforementioned limitations through a two-stage MILP framework that:

\begin{itemize}
    \item Embeds detailed industrial constraints—including binary operation decisions, inventory dynamics, and battery cycling behavior—within an integrated optimization model.
    \item Quantifies the cost of deviating from a cost-optimal baseline and compares it to expected revenues in Spain’s tertiary regulation (mFRR) market.
    \item Demonstrates the techno-economic feasibility of demand-side flexibility using actual data from a Portland cement plant and electricity market records from 2023.
\end{itemize}

In doing so, the paper offers an actionable methodology for integrating industrial flexibility into power system operations.

\medskip

The next section introduces the industrial case study and market context that underpin the proposed methodology.

\section{Industrial Context and Problem Description}
\label{sec:prob_stat}

This section situates the proposed optimization framework within the operational context of a real Portland cement manufacturing facility in Spain. It outlines the production process, identifies subsystems relevant to energy flexibility, introduces the energy assets under consideration, and clarifies the optimization objectives and market setting. Together, these elements define the scope and practical relevance of the study.

\subsection{Overview of the Cement Manufacturing Process}
\label{subsec:process_overview}

\begin{figure}[h]
\centering
\includegraphics[width=.9\columnwidth]{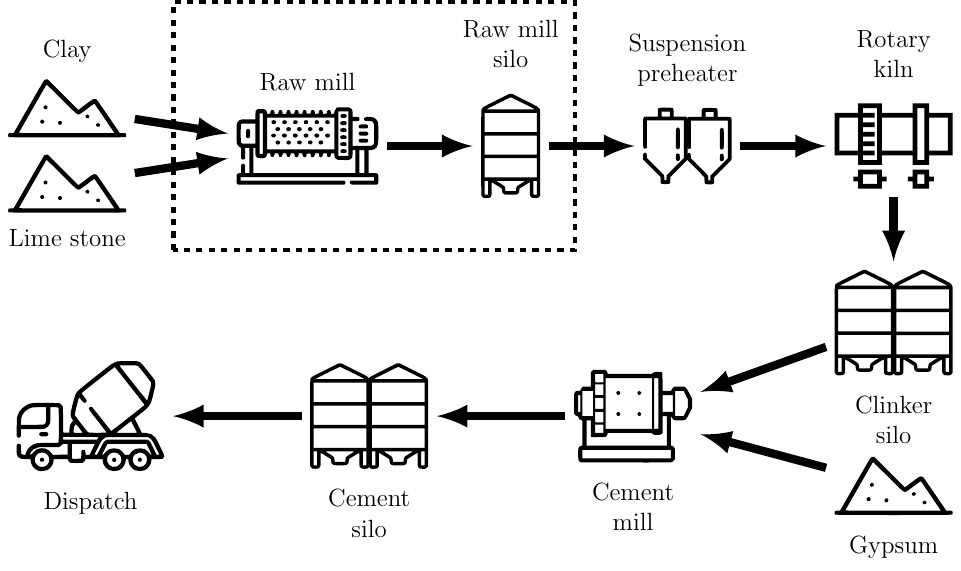}
\caption{Schematic of Portland cement production. The highlighted raw meal preparation stage is decoupled from kiln operation and offers flexibility potential.}
\label{fig:CementProd}
\end{figure}

Portland cement production is a highly energy-intensive process comprising two main stages: (i) \textit{clinker production} and (ii) \textit{cement grinding}. As illustrated in Figure~\ref{fig:CementProd}, the clinker production stage involves the extraction and processing of raw materials (primarily limestone and clay), which are crushed, dried, and finely ground into a homogeneous raw meal. This meal is temporarily stored in silos before being fed to the kiln, where it is sintered at temperatures between \SIrange{1400}{1500}{\degreeCelsius}~\cite{thermal_bal_cement} to form clinker. The hot clinker is then cooled and stored.

In the second stage, the clinker is mixed with gypsum and other additives and ground into cement with the desired physical and chemical properties. The final product is stored in silos for subsequent distribution.

\subsection{Identifying Flexibility Potential}

Among the various sub-processes, the raw meal preparation stage—highlighted in Figure~\ref{fig:CementProd}—offers the greatest potential for operational flexibility. This subsystem accounts for approximately \SIrange{20}{30}{\percent} of total site electricity consumption and is decoupled from the continuous kiln operation via an intermediate storage silo.

In contrast, the kiln must operate continuously under stringent thermal conditions, leaving limited flexibility. Similarly, the cement grinding phase is constrained by limited storage capacity and volatile product demand, making it less suitable for short-term rescheduling.

\subsection{Energy Assets and Infrastructure Considered}

This study models the following key infrastructure components within the raw meal preparation stage:

\begin{itemize}
    \item \textbf{Raw Mill:} A single mill with a rated power of \SI{6}{\mega\watt}, operating in binary on/off states. It represents the facility’s principal flexible electrical load and is part of the existing infrastructure.
    \item \textbf{Raw Meal Silo:} A large silo with a storage capacity of \SI{15,000}{\tonne}, enabling temporal decoupling between raw meal production and kiln feeding.
    \item \textbf{Prospective PV and Battery Energy Storage Systems (BESS):} Several hypothetical configurations of on-site photovoltaic generation and battery storage are introduced for techno-economic evaluation. These range from \SI{1}{\mega\watt} to \SI{6}{\mega\watt} (or \SI{6}{\mega\watt\hour}), and although not currently installed, they are assessed to explore their potential for enhancing operational flexibility and reducing electricity costs.
\end{itemize}

Together, these components support a comprehensive evaluation of strategies combining load scheduling, storage dispatch, and renewable energy integration.

\subsection{Optimization Goals and Scope}

The analysis adopts a two-stage optimization framework implemented over a rolling time horizon ranging from 24 to 168 hours, with an hourly resolution. The primary decision variables include:

\begin{enumerate}
    \item \textbf{Mill Operation:} Binary scheduling of the raw mill’s on/off status.
    \item \textbf{Silo Inventory Management:} Balancing of raw meal stocks to buffer production and demand.
    \item \textbf{PV and BESS Dispatch:} Hourly control of battery charging/discharging and allocation of PV generation between self-consumption and grid import.
\end{enumerate}

These decisions are evaluated in two consecutive stages:

\begin{description}
    \item[Stage 1 — Cost Minimization:] Determines a baseline production schedule that minimizes electricity costs using day-ahead prices, while satisfying all technical and operational constraints. This stage represents standard energy-aware scheduling from a price-taker perspective.

    \item[Stage 2 — Flexibility Evaluation:] Starting from the baseline, this stage quantifies the cost of offering load flexibility (e.g., $\pm \Delta P_\tau$~MW at time $\tau$) to Spain’s tertiary regulation market (mFRR). It identifies feasible deviations that preserve production reliability and estimates the net economic gain from balancing market revenues relative to the additional cost incurred.
\end{description}

This dual-stage formulation enables the plant to remain operationally robust while dynamically capturing short-term market opportunities.

\subsection{Balancing Market Context}

Spain’s electricity market is structured into several layers, including long-term forward contracts, the day-ahead and intra-day spot markets, and ancillary services markets. Balancing services are further divided into three categories: \textit{primary} (frequency containment reserve, FCR), \textit{secondary} (automatic frequency restoration reserve, aFRR), and \textit{tertiary} (manual frequency restoration reserve, mFRR) regulation~\cite{REE}.

This study targets the mFRR segment, which is particularly well-suited to the response capabilities of cement plant subsystems. The mFRR market allows manual activations with lead times of up to 15 minutes and durations of up to two hours, offering a practical window for rescheduling or shifting industrial loads.

By aligning the physical flexibility of the raw mill with the operational requirements of the mFRR market, the proposed methodology establishes a viable path for industrial facilities to participate in electricity markets beyond traditional procurement—enhancing both economic performance and grid reliability.

\medskip

With the industrial and market context established, the next section formalizes the optimization models used in the two-stage methodology.

\section{Methodology}
\label{sec:methodology}

This section presents the methodological framework developed to identify, quantify, and monetize short-term flexibility in the operation of a cement plant. The approach is structured into two sequential stages. Stage~1 computes an optimal production schedule that minimizes electricity procurement costs under technical and operational constraints. Stage~2 assesses the economic viability of deviating from this baseline schedule to provide flexibility services in the electricity balancing market. Both stages are formulated as mixed-integer linear programming (MILP) problems.

\subsection{Stage~1 — Optimal Production Scheduling}
\label{subsec:stage1}

Stage~1 determines a cost-optimal baseline schedule for the raw milling subsystem of the plant. This schedule serves as the reference for evaluating market-responsive flexibility in Stage~2. All inputs—including day-ahead electricity prices, photovoltaic (PV) generation forecasts, and process demand—are treated as deterministic. Although stochastic formulations are relevant, they are considered outside the scope of this study.

The MILP is solved over a rolling planning horizon with hourly resolution, allowing for dynamic adjustments in response to updated forecasts or operational data.

\paragraph{System Scope}

\begin{figure}[h]
\centering
\includegraphics[width=.7\columnwidth]{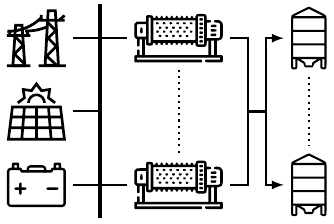}
\caption{Schematic of the modeled production subsystem, including raw mills, silos, PV generation, BESS, and grid connection. The model supports scalability to multi-unit scenarios.}
\label{fig:Plant}
\end{figure}

Figure~\ref{fig:Plant} illustrates the modeled subsystem, which includes raw mills, silos, PV installations, battery energy storage systems (BESS), and a connection to the public grid. Although the case study focuses on a single unit of each component, the model is general and scalable to multi-unit configurations. The optimization goal is to satisfy kiln demand \( D_t \) at each time step while minimizing electricity purchases from the grid.

\subsubsection*{Model Components}

\textbf{Index Sets}
\begin{itemize}
    \item $\mathcal{T} = \{1, \dots, N_T\}$: Time periods in the planning horizon (e.g., hourly), with step size $\Delta t = \SI{1}{h}$.
    \item $\mathcal{K} = \{1, \dots, N_K\}$: Set of raw milling units.
\end{itemize}

\textbf{Key Parameters}
\begin{itemize}
    \item $\pi_t$ [\euro{}/MWh]: Forecasted day-ahead electricity price.
    \item $P_{\mathrm{PV},t}$ [MW]: Forecasted PV generation at time $t$.
    \item $P_k$ [MW], $\Pi_k$ [t/h]: Rated power and throughput of mill $k$.
    \item $I_{k,\min}, I_{k,\max}$ [t]: Inventory bounds for silo connected to mill $k$.
    \item $C_{\max}$ [MWh]: BESS energy capacity.
    \item $P_C^{\max}, P_D^{\max}$ [MW]: BESS maximum charging and discharging rates.
    \item $M_k^{\mathrm{ON}}, M_k^{\mathrm{OFF}}$ [h]: Minimum on/off durations for mill $k$.
    \item $P_b^{\max}$ [MW]: Maximum grid import capacity.
\end{itemize}

\textbf{Decision Variables}
\begin{align*}
Y_{k,t} &\in \{0,1\} && \text{On/off status of mill $k$} \\
I_{k,t} &\ge 0 && \text{Inventory in silo $k$} \\
P_{b,t}, P_{s,t} &\ge 0 && \text{Grid import/export power} \\
P_{C,t}, P_{D,t} &\ge 0 && \text{BESS charge/discharge power} \\
\mathrm{SOC}_t &\ge 0 && \text{State of charge of BESS}
\end{align*}

\subsubsection*{Objective Function}

The optimization objective is to minimize electricity procurement costs from the grid:
\begin{equation} \label{eq:stage1_obj}
\min\; \Phi = \sum_{t \in \mathcal{T}} \pi_t \cdot P_{b,t} \cdot \Delta t.
\end{equation}

Only grid purchases contribute to the cost function. PV generation and battery discharges are considered free; grid exports do not yield revenue in this formulation.

\subsubsection*{Constraints}

Unless otherwise noted, all constraints apply to each time step \( t \in \mathcal{T} \) and each mill \( k \in \mathcal{K} \).

\textbf{1) Silo material balance}
\begin{equation} \label{eq:mat_bal}
I_{k,t} = I_{k,t-1} + \Pi_k Y_{k,t} - D_{k,t}.
\end{equation}

This equation governs inventory evolution. When mill \( k \) is active, it produces material at a rate \( \Pi_k \), which is added to the silo. At the same time, material is withdrawn from the silo to meet the process demand \( D_{k,t} \).

\textbf{2) Power balance}
\begin{equation} \label{eq:pow_bal}
P_{b,t} + P_{D,t} + P_{\mathrm{PV},t} = P_{s,t} + P_{C,t} + \sum_{k \in \mathcal{K}} P_k Y_{k,t}.
\end{equation}

This constraint ensures that total power supply—comprising grid import, battery discharge, and PV generation—matches consumption, which includes grid export, battery charging, and mill operation.

\textbf{3) Silo inventory limits}
\begin{equation} \label{eq:silo_bounds}
I_{k,\min} \le I_{k,t} \le I_{k,\max}.
\end{equation}

Each silo must operate within its defined storage capacity bounds to preserve material buffering and avoid overflow or depletion.

\textbf{4) Demand coverage}
\begin{equation} \label{eq:demand_cover}
\sum_{k \in \mathcal{K}} I_{k,t} \ge D_t.
\end{equation}

This constraint guarantees that the cumulative inventory across all silos is sufficient to meet the hourly kiln demand \( D_t \). If needed, the total demand can be disaggregated across mills as \( D_t = \sum_k D_{k,t} \).

\textbf{5) Battery dynamics and operating limits}
\begin{subequations} \label{eq:batt}
\begin{align}
\mathrm{SOC}_t &= \mathrm{SOC}_{t-1} + (P_{C,t} - P_{D,t}) \cdot \Delta t, \label{eq:batt_soc} \\
C_{\max}(1 - \mathrm{DoD}) &\le \mathrm{SOC}_t \le C_{\max}, \label{eq:batt_bounds} \\
0 &\le P_{C,t} \le P_C^{\max}, \quad 0 \le P_{D,t} \le P_D^{\max}. \label{eq:batt_power}
\end{align}
\end{subequations}

The state of charge (SOC) evolves based on the net energy flow. Constraint \eqref{eq:batt_bounds} ensures that the SOC remains within the usable energy window, considering a minimum depth of discharge (DoD). Power limits constrain instantaneous charging and discharging capabilities. The initial condition is:
\[
\mathrm{SOC}_0 = 0.5 \cdot C_{\max}.
\]

\textbf{6) Grid import limit}
\begin{equation} \label{eq:grid_cap}
0 \le P_{b,t} \le P_b^{\max}.
\end{equation}

This constraint enforces the grid connection's maximum import capacity, preventing overload.

\textbf{7) Minimum mill uptime and downtime}
\begin{subequations} \label{eq:min_times}
\begin{align}
(Y_{k,t} - Y_{k,t-1}) \cdot M_k^{\mathrm{ON}} &\le \sum_{j=0}^{M_k^{\mathrm{ON}} - 1} Y_{k,t+j}, \\
\sum_{j=0}^{M_k^{\mathrm{OFF}} - 1} Y_{k,t+j} &\le (1 + Y_{k,t-1} - Y_{k,t}) \cdot M_k^{\mathrm{OFF}}.
\end{align}
\end{subequations}

These constraints ensure that once a mill is turned on (or off), it must remain in that state for a minimum number of consecutive hours, reducing wear and tear and reflecting operational stability requirements.

\paragraph{Solution and Interpretation}

The MILP defined by \eqref{eq:stage1_obj}–\eqref{eq:min_times} is solved using commercial (e.g., Gurobi) or open-source (e.g., SCIP) solvers. The output is a cost-optimal baseline schedule:
\[
\{Y_{k,t}^*, I_{k,t}^*, P_{b,t}^*, P_{s,t}^*, \mathrm{SOC}_t^*, \ldots\}
\]
with associated total electricity cost \( \Phi^* \). This solution serves as input to the flexibility optimization in Stage~2.

Table~\ref{tab:notation} summarizes all key symbols used in the Stage~1 formulation.

\begin{table*}[ht]
\centering
\caption{Key notation used in the Stage~1 optimization model.}
\label{tab:notation}
\begin{tabular*}{\textwidth}{@{\extracolsep{\fill}}lll}
\toprule
\textbf{Symbol} & \textbf{Description} & \textbf{Unit / Type} \\
\midrule
$\mathcal{T}$ & Time periods in planning horizon & index set \\
$\mathcal{K}$ & Raw milling units & index set \\
$Y_{k,t}$ & On/off status of mill $k$ & binary \\
$I_{k,t}$ & Inventory in silo $k$ & [t] \\
$P_{b,t}, P_{s,t}$ & Grid import/export power & [MW] \\
$P_{C,t}, P_{D,t}$ & BESS charging/discharging power & [MW] \\
$\mathrm{SOC}_t$ & BESS state of charge & [MWh] \\
$\pi_t$ & Day-ahead electricity price & [€/MWh] \\
$P_{\mathrm{PV},t}$ & PV generation & [MW] \\
$P_k$, $\Pi_k$ & Power and throughput of mill $k$ & [MW], [t/h] \\
$I_{k,\min}, I_{k,\max}$ & Silo inventory limits & [t] \\
$C_{\max}$ & Battery capacity & [MWh] \\
$P_C^{\max}, P_D^{\max}$ & BESS power limits & [MW] \\
$M_k^{\mathrm{ON}}, M_k^{\mathrm{OFF}}$ & Min. uptime/downtime & [h] \\
$P_b^{\max}$ & Max grid import capacity & [MW] \\
$D_t$, $D_{k,t}$ & Total and per-silo process demand & [t/h] \\
$\Phi$ & Total electricity cost & [€] \\
\bottomrule
\end{tabular*}
\end{table*}

\subsection{Stage 2 — Flexibility Evaluation}
\label{sec:flex_prog}

The second stage evaluates whether the cement plant can profitably deviate from its cost-optimal baseline schedule in order to provide short-term flexibility in the Spanish tertiary regulation market (manual Frequency Restoration Reserve, mFRR). This is achieved by simulating enforced load deviations and quantifying the tradeoff between balancing market revenue and the cost of deviating from the baseline.

Table~\ref{tab:flex_symbols} summarizes the notation specific to this flexibility evaluation model.

\begin{table*}[ht]
\centering\small
\caption{Notation specific to the flexibility evaluation model (Stage 2).}
\label{tab:flex_symbols}
\begin{tabular}{lll}
\toprule
\textbf{Symbol} & \textbf{Description} & \textbf{Unit / Type} \\
\midrule
$\tau$ & Activation hour for flexibility provision & [h] (index) \\
$\Delta P_\tau$ & Enforced power deviation at time $\tau$ & [MW] \\
$P_{b,\tau}^*$ & Baseline grid import at time $\tau$ & [MW] \\
$I_{k,\tau}^*$ & Baseline inventory for mill $k$ at time $\tau$ & [t] \\
$\Phi^*$ & Cost of the baseline schedule (Stage 1) & [€] \\
$\Phi^\dagger$ & Cost of the flexibility-adjusted schedule & [€] \\
$\Delta \Phi$ & Flexibility cost: $\Phi^\dagger - \Phi^*$ & [€] \\
$\lambda_\tau^{\text{mFRR}\,+}$ & Upward regulation price (load reduction) & [€/MWh] \\
$\lambda_\tau^{\text{mFRR}\,-}$ & Downward regulation price (load increase) & [€/MWh] \\
$\lambda_\tau^{\text{spread}}$ & Minimum break-even price spread & [€/MWh] \\
$E_b^*$ & Total energy purchased under baseline & [MWh] \\
$\varepsilon^-, \varepsilon^+$ & Energy deviation tolerances & [0–1] \\
\bottomrule
\end{tabular}
\end{table*}

\paragraph{Conceptual Overview}

Flexibility is modeled as a forced, single-period deviation at a specific activation time \( \tau \in \mathcal{T}_{\mathrm{act}} \). A power shift \( \Delta P_\tau \) is imposed: positive for upward regulation (load reduction), negative for downward regulation (load increase). The optimization problem is then re-solved for the remaining time horizon, subject to technical feasibility and energy neutrality constraints.

\paragraph{Modified Constraints}

The Stage 2 model inherits all structural constraints from Stage 1, with the following modifications to reflect the activation of flexibility:

\begin{itemize}
    \item \textbf{Frozen pre-activation decisions:} For all periods \( t < \tau \), key variables are fixed to their baseline values:
    \begin{align*}
    Y_{k,t} &= Y_{k,t}^*, \quad I_{k,t} = I_{k,t}^*, \\
    P_{b,t} &= P_{b,t}^*, \quad \mathrm{SOC}_t = \mathrm{SOC}_t^*.
    \end{align*}

    This ensures that the past operational history is preserved and only post-\( \tau \) behavior is re-optimized.

    \item \textbf{Enforced deviation at activation:} At time \( t = \tau \), the grid import is modified relative to the baseline:
    \begin{equation}
    P_{b,\tau} = P_{b,\tau}^* + \Delta P_\tau.
    \end{equation}

    \item \textbf{Energy neutrality:} Over the full horizon, total energy imported from the grid must remain close to the baseline value \( E_b^* \), within symmetric or asymmetric tolerance bounds:
    \begin{equation}
    (1 - \varepsilon^-) E_b^* \le \sum_{t \in \mathcal{T}} P_{b,t} \cdot \Delta t \le (1 + \varepsilon^+) E_b^*.
    \end{equation}
    This prevents unrealistic net load shifts across the planning period and ensures alignment with physical energy use.
\end{itemize}

Unlike pre-\( \tau \) variables, post-\( \tau \) mill operating hours and inventory trajectories are allowed to adjust in order to recover from the deviation, as long as all constraints remain satisfied.

\paragraph{Flexibility Cost and Profitability}

The cost of the re-optimized schedule under deviation is denoted \( \Phi^\dagger \), and the flexibility cost is:
\begin{equation}
\Delta \Phi(\Delta P_\tau, \tau) = \Phi^\dagger - \Phi^*.
\end{equation}

A flexibility offer is profitable if the revenue earned in the mFRR market exceeds the incurred cost:
\begin{equation}
|\Delta P_\tau| \cdot \Delta t \cdot \lambda_\tau^{\text{mFRR}\,\pm} > \Delta \Phi(\Delta P_\tau, \tau),
\end{equation}
where \( \lambda_\tau^{\text{mFRR}\,\pm} \) is the market clearing price for upward ( \( \Delta P_\tau > 0 \) ) or downward ( \( \Delta P_\tau < 0 \) ) reserve.

The corresponding break-even spread is given by:
\begin{equation}
\lambda_\tau^{\text{spread}} = \frac{\Delta \Phi(\Delta P_\tau, \tau)}{|\Delta P_\tau| \cdot \Delta t}.
\end{equation}

This metric provides a useful benchmark: market prices above this spread indicate profitable transactions.

\paragraph{Implementation Workflow}

The flexibility assessment is implemented over a rolling horizon and multiple scenarios, following this process:

\begin{enumerate}
    \item Select candidate activation times \( \tau \) and deviation values \( \Delta P_\tau \) for simulation.
    \item For each \( (\tau, \Delta P_\tau) \) pair:
    \begin{itemize}
        \item Fix pre-\( \tau \) variables to baseline values.
        \item Enforce the deviation at time \( \tau \).
        \item Re-solve the MILP for the remaining horizon.
    \end{itemize}
    \item Compute the flexibility cost \( \Delta \Phi(\Delta P_\tau, \tau) \).
    \item Compare the result with expected mFRR prices to determine profitability.
\end{enumerate}

\bigskip

This second stage complements the baseline scheduling developed in Stage~1 by enabling economically rational and technically feasible participation in balancing markets. The following section applies the full methodology to a real-world cement manufacturing facility in Spain.

\section{Case Study — Application to a Spanish Portland‑Cement Plant}
\label{Case_study_main_chap}

This section demonstrates the practical application of the proposed methodology to a medium-sized Portland cement plant in Spain. The analysis draws on real operational data, electricity market prices, and technical parameters provided by the plant operator and project collaborators. The objective is to assess the techno-economic potential of flexibility using the two-stage optimization framework outlined in Section~\ref{sec:methodology}.

The section is organized as follows. First, we describe the energy-relevant infrastructure of the facility. Then, we introduce the input data and scenario matrix, along with the rationale for selecting two representative months from 2023. Finally, we present the simulation design and evaluation approach.

\subsection{Plant Description}

The facility operates a single raw milling line that consumes approximately \SI{6}{MW} of electricity when running at full capacity. At this load, the mill produces raw meal at a rate of \SI{360}{t/h}. Milled material is stored in a silo with a total capacity of \SI{15,000}{t}, and a minimum inventory threshold of \SI{9,000}{t} is maintained to ensure at least six hours of uninterrupted kiln feeding during mill downtime.

While the plant currently relies exclusively on grid electricity, it is actively evaluating the integration of on-site generation and storage technologies to enhance operational flexibility and reduce energy costs. Specifically, two technology types are considered:

\begin{itemize}
  \item \textbf{Photovoltaic (PV) array:} Ground-mounted, with installed capacities ranging from 1 to \SI{6}{MWp}.
  \item \textbf{Battery Energy Storage System (BESS):} Lithium-ion systems with capacities from 1 to \SI{6}{MWh}, rated at 1C with an 80\% usable depth-of-discharge.
\end{itemize}

All scenarios respect the plant’s contracted maximum grid import limit of \SI{21}{MW}.

\subsection{Data Sources}
\label{subsec:data_sources}

The simulation relies on real-world input data from the following sources:

\begin{itemize}
  \item \textbf{Day-ahead prices (\( \pi_t \)):} Hourly electricity prices for April and June 2023 from the OMIE market. The Stage~1 scheduling model uses forecasted prices generated by Fortia’s proprietary algorithm~\cite{sebastian2023adaptive}.
  \item \textbf{Balancing market prices (\( \lambda_t^{\pm} \)):} Upward and downward mFRR clearing prices published by Red Eléctrica de España (REE).
  \item \textbf{PV generation (\( P_{\mathrm{PV},t} \)):} Estimated using clear-sky irradiance data from the PVGIS database, based on the plant’s geographical coordinates.
  \item \textbf{Process demand and constraints:} The kiln requires a relatively stable raw meal supply of \SI{240}{t/h}, with values and constraints confirmed by plant personnel.
\end{itemize}

All time series are synchronized to Central European Time (UTC+1) to avoid discontinuities due to daylight saving changes.

\subsection{Scenario Matrix}
\label{subsec:scenarios}

A total of 19 technology configurations are simulated, including the current setup without PV or BESS. Each scenario is labeled \texttt{MXY}, where \texttt{X} denotes photovoltaic (PV) capacity in megawatts peak (MWp), and \texttt{Y} denotes battery energy storage system (BESS) capacity in megawatt-hours (MWh). The matrix encompasses PV-only, BESS-only, and combined PV+BESS configurations.

To limit model complexity while capturing scale effects, only matched PV+BESS combinations (\texttt{M11}, \texttt{M22}, ..., \texttt{M66}) were simulated, representing balanced deployments of generation and storage. 

The full set of simulated configurations is summarized in Table~\ref{tab:scenarios}, which also describes the purpose and classification of each scenario type.

\begin{table*}[!htb]
\centering
\caption{Technology configurations simulated in the case study. Scenario code \texttt{MXY} indicates PV capacity \( X \) (MWp) and BESS capacity \( Y \) (MWh).}
\label{tab:scenarios}
\scriptsize 
\setlength{\tabcolsep}{4pt} 
\begin{tabular*}{\textwidth}{@{\extracolsep{\fill}}lllll@{}}
\toprule
\textbf{Scenario Code(s)} & \textbf{PV Capacity (MWp)} & \textbf{BESS Capacity (MWh)} & \textbf{Configuration Type} & \textbf{Purpose} \\
\midrule
\texttt{M00} & 0 & 0 & No PV, no BESS & Reference case \\
\midrule
\texttt{M10--M60} & 1--6 & 0 & PV only & RES scale-up analysis \\
\texttt{M01--M06} & 0 & 1--6 & BESS only & Storage-only evaluation \\
\texttt{M11}, \texttt{M22}, ..., \texttt{M66} & 1--6 & 1--6 (diagonal only) & PV + BESS & Balanced combined strategies \\
\bottomrule
\end{tabular*}
\end{table*}

\paragraph{Why April and June 2023?}

\begin{figure}[h]
\centering
\includegraphics[width=.45\textwidth]{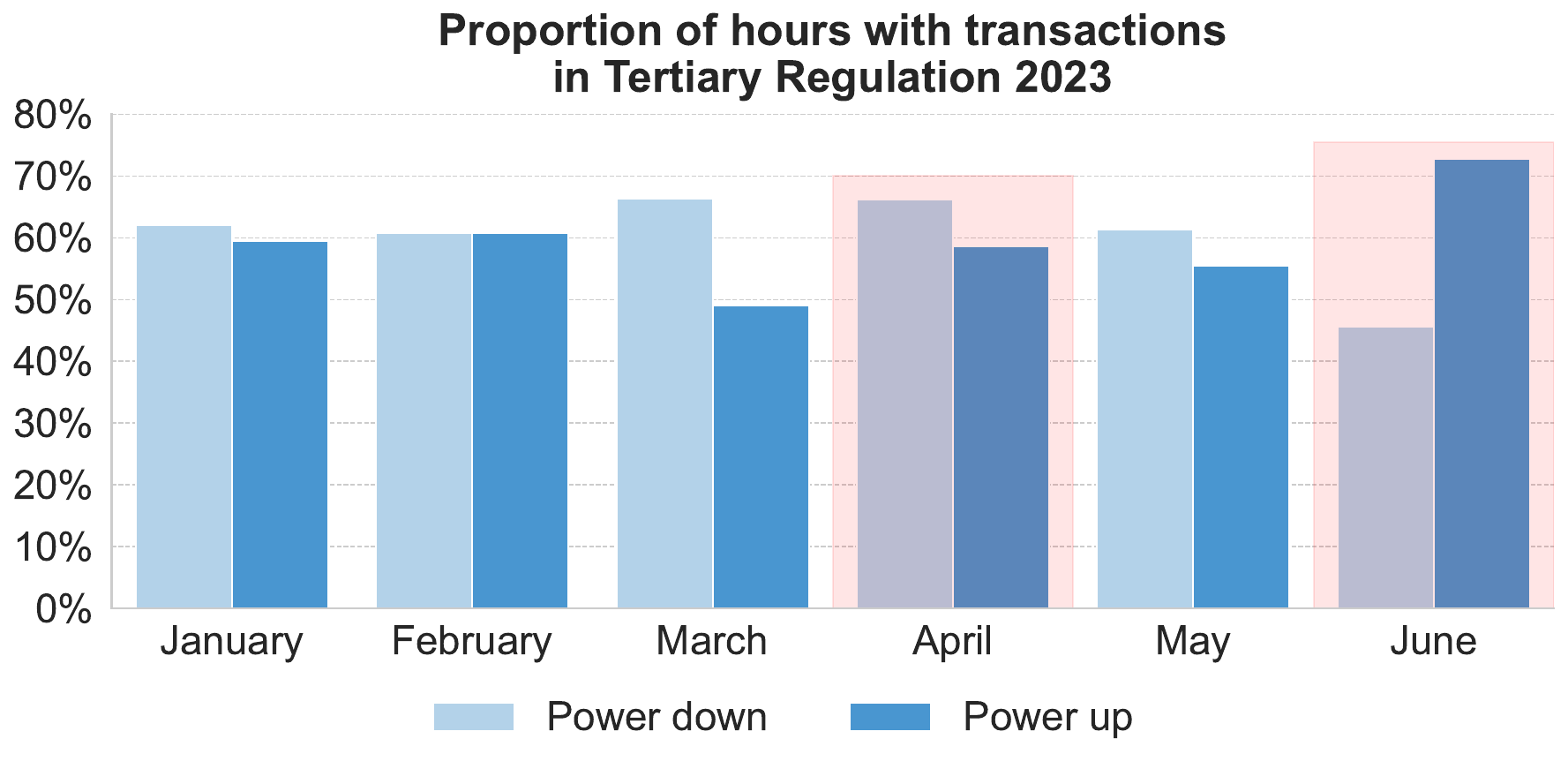}
\includegraphics[width=.45\textwidth]{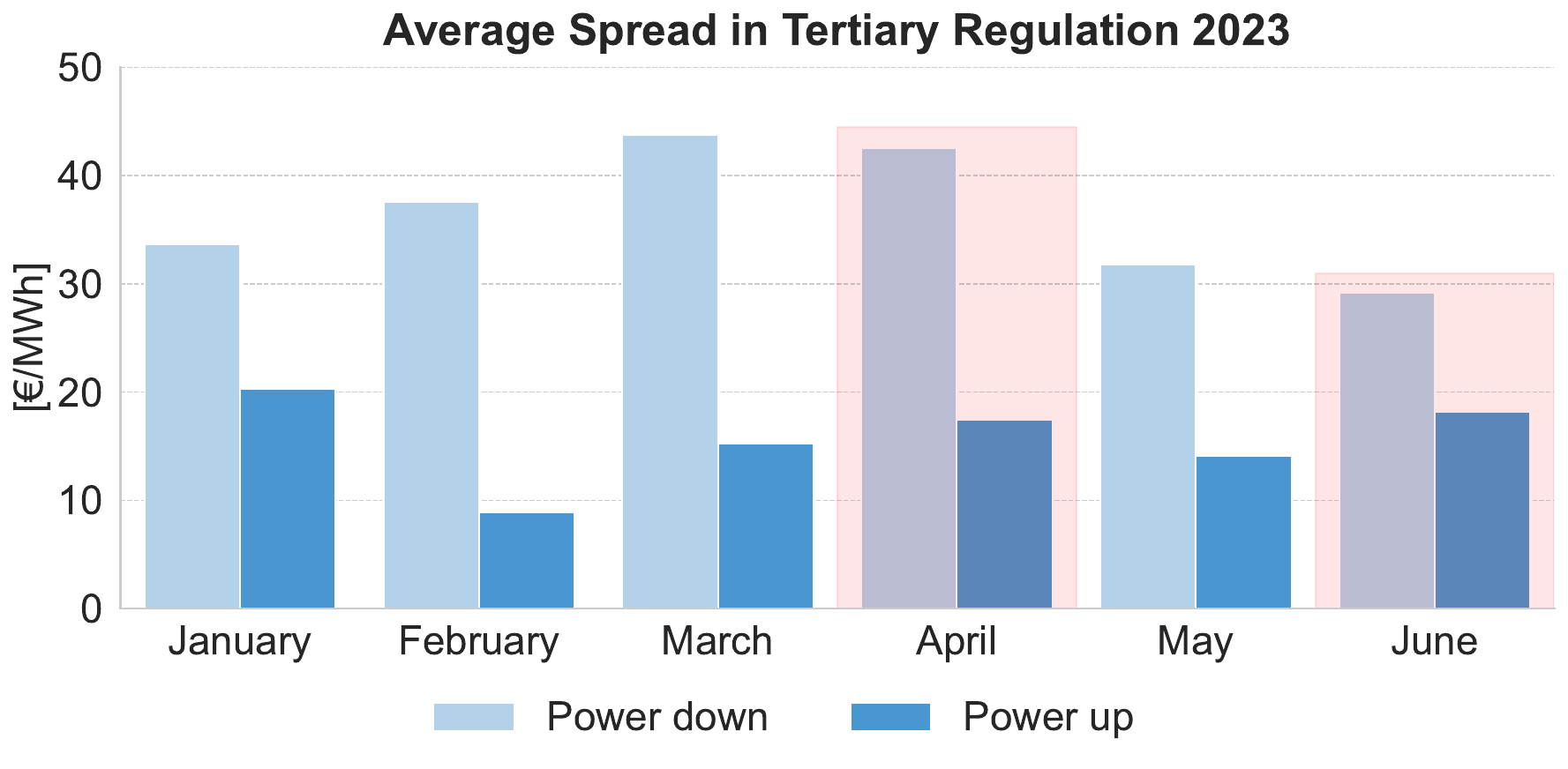}
\caption{Top: Share of hours with activation in the tertiary regulation (mFRR) market in 2023, disaggregated by direction (upward = load reduction, downward = load increase). Bottom: Corresponding average price spreads (\euro/MWh) between day-ahead and balancing markets, reflecting the financial incentive for flexible participation. April shows high activation and high spread levels; June shows reduced opportunities dominated by downward regulation.}
\label{fig:tert_graph}
\end{figure}

April and June represent contrasting market conditions that are well-suited for evaluating the performance and robustness of the proposed flexibility strategy. Figure~\ref{fig:tert_graph} illustrates the key differences across these two months:

\begin{itemize}
  \item \textbf{April 2023:} Exhibited a high share of hours with mFRR activations in both upward (load reduction) and downward (load increase) directions. It also recorded one of the highest average price spreads and relatively balanced activation patterns—indicating strong technical and economic signals for flexible participation.

  \item \textbf{June 2023:} Characterized by a sharp decline in upward activations and a dominance of downward regulation calls. Average price spreads were lower but more symmetric between directions, reducing the overall profitability of flexibility actions and making it a more challenging environment for value extraction.
\end{itemize}

Together, these two months enable a robust evaluation across both favorable and asymmetric market scenarios.

\subsection{Simulation Design}
\label{subsec:sim_design}

\paragraph{Planning horizon and rolling strategy.}

Each simulation spans a 30-day period in either April or June 2023, using hourly resolution. A rolling-horizon implementation is applied to reflect realistic plant operations and market participation:

\begin{enumerate}
  \item At midnight each day, Stage~1 computes a 7-day (168-hour) baseline schedule based on updated forecasts.
  \item Stage~2 evaluates flexibility transactions for the first 24 hours of that window, testing \( \Delta P_\tau \in \{-6, +6\} \) MW.
  \item The planning horizon then advances by one day, and the procedure repeats.
\end{enumerate}

\paragraph{Solver and computation time.}

All models are implemented in \texttt{PySCIPOpt}~\cite{SCIP_solver_Python, SCIP_solver} and solved using \textsc{SCIP~8.0.3} with default settings. Average runtimes were:

\begin{itemize}
  \item \textbf{Stage 1:} \SI{1.9}{s} per 7-day scheduling problem.
  \item \textbf{Stage 2:} \SI{0.15}{s} per flexibility evaluation (per \( \tau \) and \( h \)).
\end{itemize}

In total, over 4,250 MILP instances were solved across all scenario-month combinations, with an aggregate runtime of approximately 3.3 hours.

\bigskip

With the case study design and assumptions established, the next section presents the results of the simulations and discusses key operational and economic insights across the tested configurations.

\section{Results and Discussion}
\label{sec:results}

This section presents the numerical results of applying the two-stage baseline-and-flexibility framework to the Spanish cement plant. It evaluates the model’s outputs across several dimensions, providing a comprehensive picture of flexibility performance, techno-economic impact, and investment viability.

The discussion is structured into five parts: (1) illustrative examples of flexibility transactions, (2) aggregated economic performance, (3) availability of flexible hours and success rates, (4) electricity cost savings from asset-driven scheduling, and (5) investment payback period estimates.

\subsection{Illustrative Flexibility Transactions}

\begin{figure}[h]
\centering
\includegraphics[width=.45\textwidth]{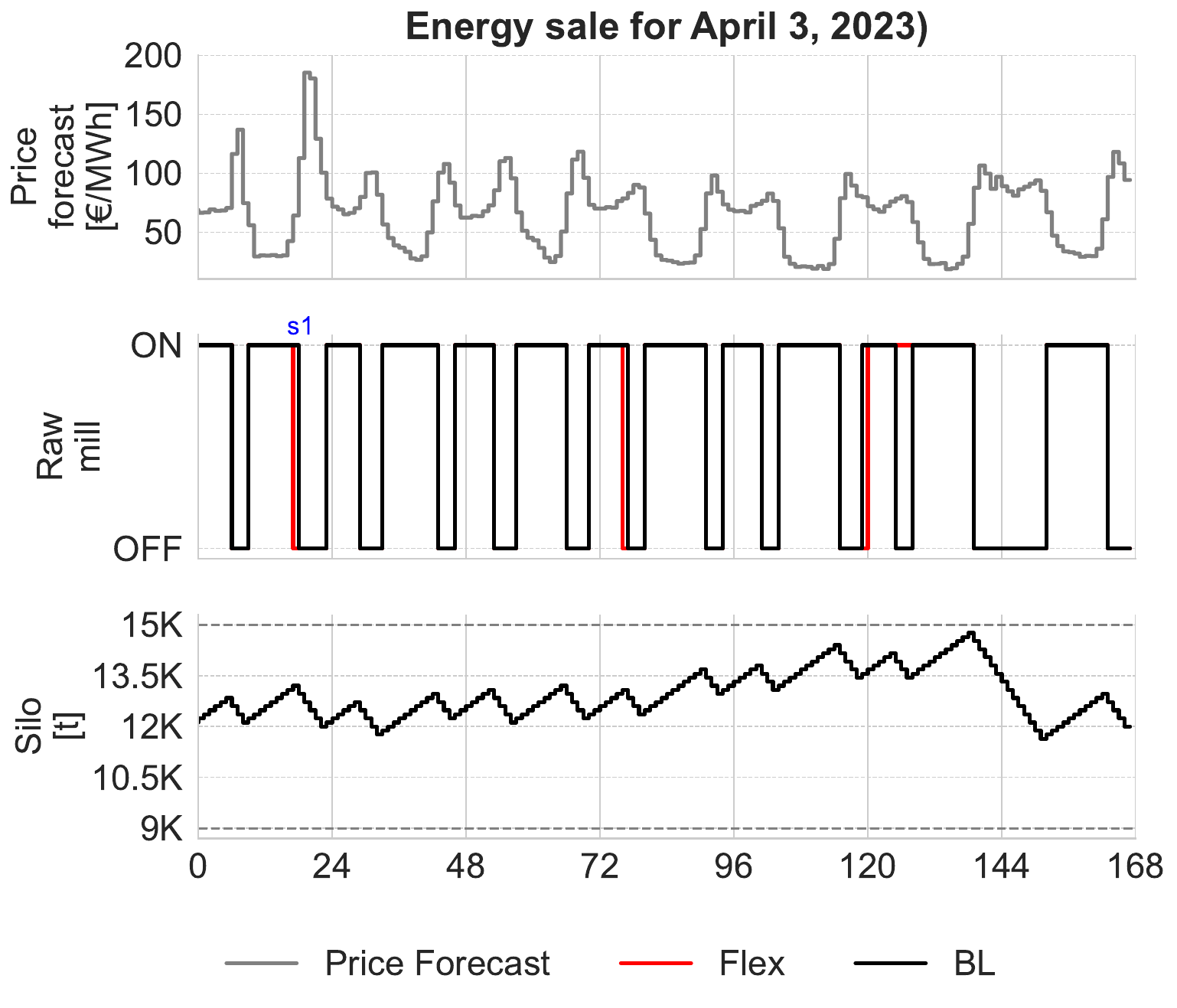}
\includegraphics[width=.45\textwidth]{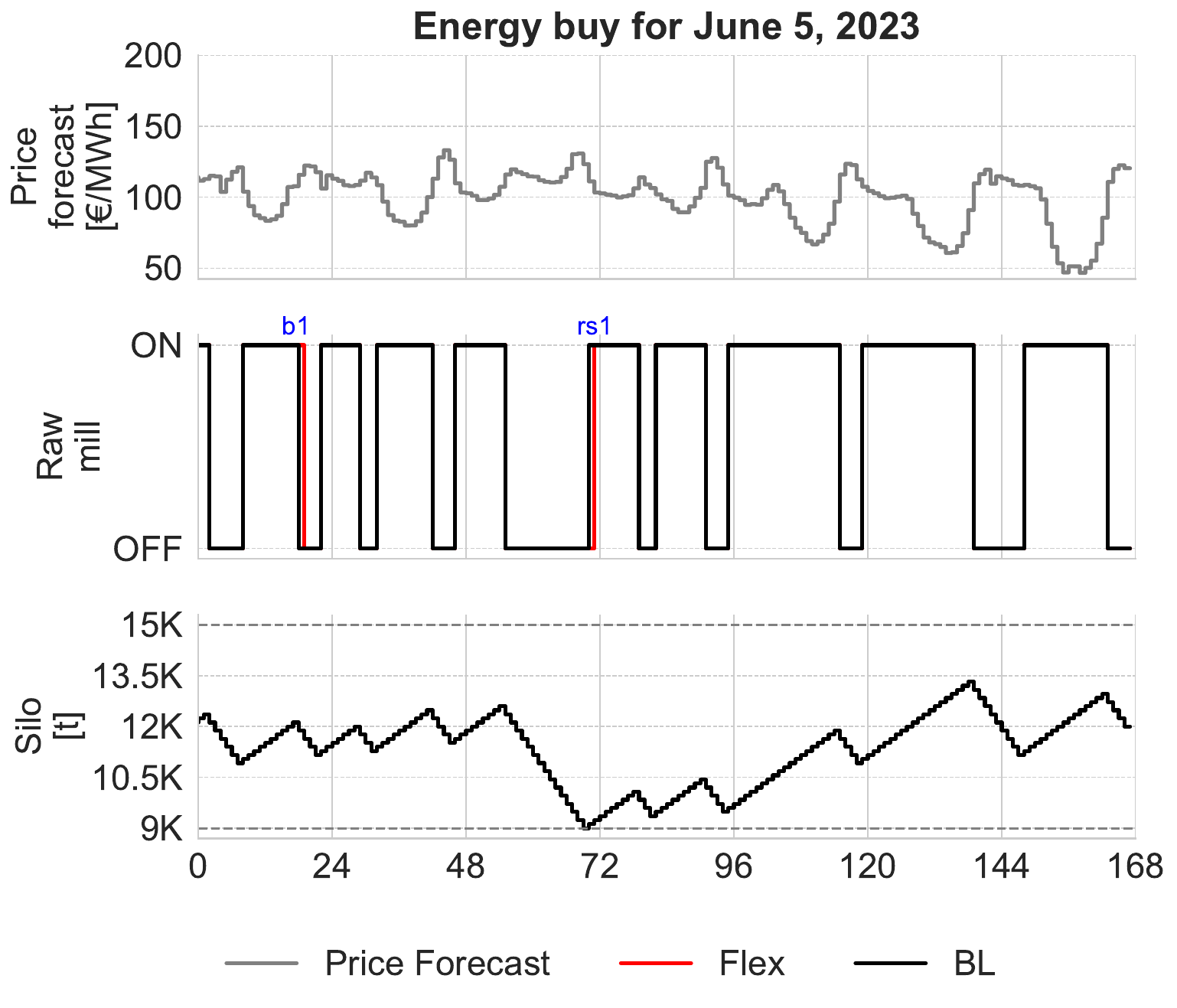}
\caption{Top: Selling electricity via load reduction (April 3, 2023). Bottom: Buying electricity via load increase (June 5, 2023). Black: baseline schedule. Red: flexibility-adjusted schedule.}
\label{fig:example_flex_fig}
\end{figure}

To demonstrate how the methodology operates in practice, Figure~\ref{fig:example_flex_fig} shows two representative examples where the model identifies profitable deviations from the baseline schedule. These reflect actual market opportunities in April and June 2023.

In the first case (top), the mill temporarily reduces load in response to a tertiary market (mFRR up) signal, capitalizing on a high price spread between the day-ahead and balancing markets. In the second (bottom), it increases consumption during a period of system overgeneration, helping the grid absorb excess supply at low cost.

Each plot consists of three layers: forecasted day-ahead prices (top), raw mill status (middle), and silo inventory (bottom). Notably, the flexibility actions are placed at the edges of production blocks, minimizing disruption to process continuity and respecting minimum uptime/downtime constraints.

In both cases, profitability hinges on whether the marginal revenue (or savings) exceeds the operational cost of deviating from the optimized schedule. This is reflected in the tables~\ref{table:example_sales_flex_tbl} and ~\ref{table:example_buy_flex_tbl}.

\begin{table*}[ht]
\centering
\caption{Profitability analysis for selling energy in the tertiary regulation market on April 3, 2023.}
\label{table:example_sales_flex_tbl}
\begin{adjustbox}{width=\textwidth}
\begin{tabular}{cccccccc}
\toprule
\textbf{Hour ($\tau$)} &
\textbf{Day-ahead price $\pi_t$} &
\textbf{mFRR price $\lambda_\tau^{\text{mFRR}\,+}$} &
\textbf{Spread $\lambda_\tau^{\text{spread}}$} &
\textbf{Power sold $\Delta P_\tau$} &
\textbf{Revenue} &
\textbf{Flex cost $\Delta\Phi$} &
\textbf{Profit} \\
{[\si{\hour}]} &
{[\si{\EUR\per{\MW\hour}}]} &
{[\si{\EUR\per{\MW\hour}}]} &
{[\si{\EUR\per{\MW\hour}}]} &
{[\si{\MW}]} &
{[\si{\EUR}]} &
{[\si{\EUR}]} &
{[\si{\EUR}]} \\
\midrule
1  & 68.97  & 84.35  & 15.38  & $-6$ & 92.25  & 45.00  & 47.25 \\
7  & 70.48  & 92.80  & 22.32  & $-6$ & 133.90 & 35.90  & 98.00 \\
11 & 55.89  & --     & --     & $-6$ & --     & 123.50 & --    \\
\rowcolor{lightgray}
19 & 64.10  & 97.28  & 33.18  & $-6$ & 199.01 & 74.20  & 124.87 \\
\bottomrule
\end{tabular}
\end{adjustbox}
\end{table*}

\begin{table*}[ht]
\centering
\caption{Profitability analysis for purchasing energy in the tertiary regulation market on June 5, 2023.}
\label{table:example_buy_flex_tbl}
\begin{adjustbox}{width=\textwidth}
\begin{tabular}{cccccccc}
\toprule
\textbf{Hour ($\tau$)} &
\textbf{Day-ahead price $\pi_t$} &
\textbf{mFRR price $\lambda_\tau^{\text{mFRR}\,-}$} &
\textbf{Spread $\lambda_\tau^{\text{spread}}$} &
\textbf{Power bought $\Delta P_\tau$} &
\textbf{Savings} &
\textbf{Flex cost $\Delta\Phi$} &
\textbf{Net benefit} \\
{[\si{\hour}]} &
{[\si{\EUR\per{\MW\hour}}]} &
{[\si{\EUR\per{\MW\hour}}]} &
{[\si{\EUR\per{\MW\hour}}]} &
{[\si{\MW}]} &
{[\si{\EUR}]} &
{[\si{\EUR}]} &
{[\si{\EUR}]} \\
\midrule
4  & 114.99 & --     & --     & 6 & --     & 23.50  & --     \\
8  & 117.89 & 61.99  & 55.90  & 6 & 354.02 & 59.50  & 294.52 \\
\rowcolor{lightgray}
20 & 115.77 & 45.56  & 70.21  & 6 & 421.24 & 28.10  & 393.14 \\
23 & 117.63 & 60.77  & 56.86  & 6 & 341.16 & 39.30  & 301.86 \\
\bottomrule
\end{tabular}
\end{adjustbox}
\end{table*}

\subsection{Aggregate Economic Performance}

\begin{figure}[h]
\centering
\includegraphics[width=.45\textwidth]{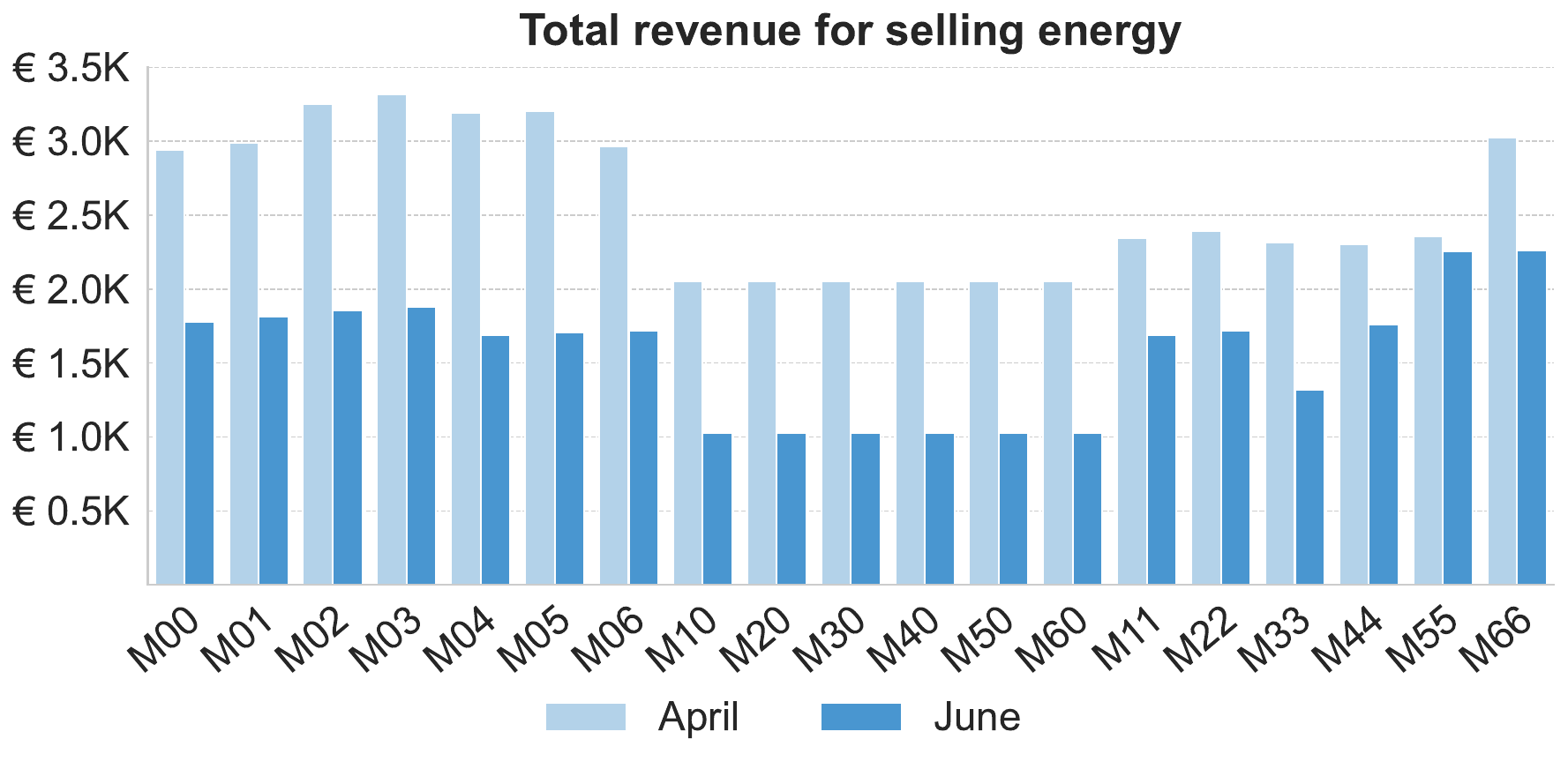}
\includegraphics[width=.45\textwidth]{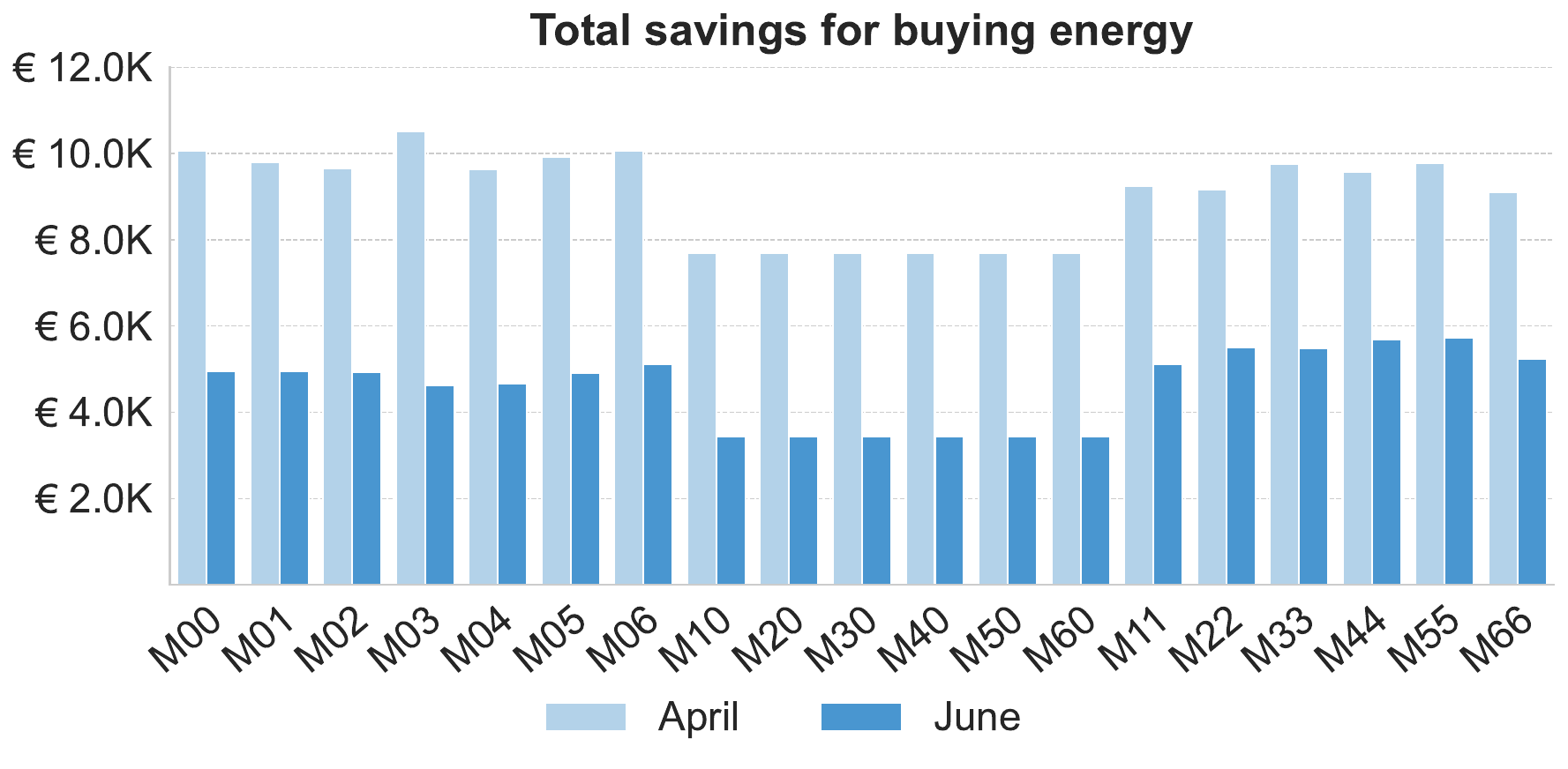}
\caption{Monthly profits from flexibility provision in the tertiary regulation market. Top: Selling (upward). Bottom: Buying (downward). Batteries significantly enhance revenues across scenarios.}
\label{fig:reven_vs_savs}
\end{figure}

Figure~\ref{fig:reven_vs_savs} aggregates monthly revenues from all profitable flexibility transactions, by scenario and market direction.

\begin{itemize}
  \item \textbf{Downward regulation} (buying electricity) is consistently more profitable than upward regulation. This reflects both the frequency of activation and the magnitude of price spreads—particularly during midday PV overgeneration.
  \item \textbf{Battery-enhanced scenarios} achieve higher revenues than PV-only cases, since BESS enables time-shifting and fast response to market events.
  \item \textbf{April outperforms June}, demonstrating the influence of market seasonality on flexibility value.
\end{itemize}

\subsection{Availability of Flexible Hours and Success Rates}

\begin{figure}[h]
\centering
\includegraphics[width=.45\textwidth]{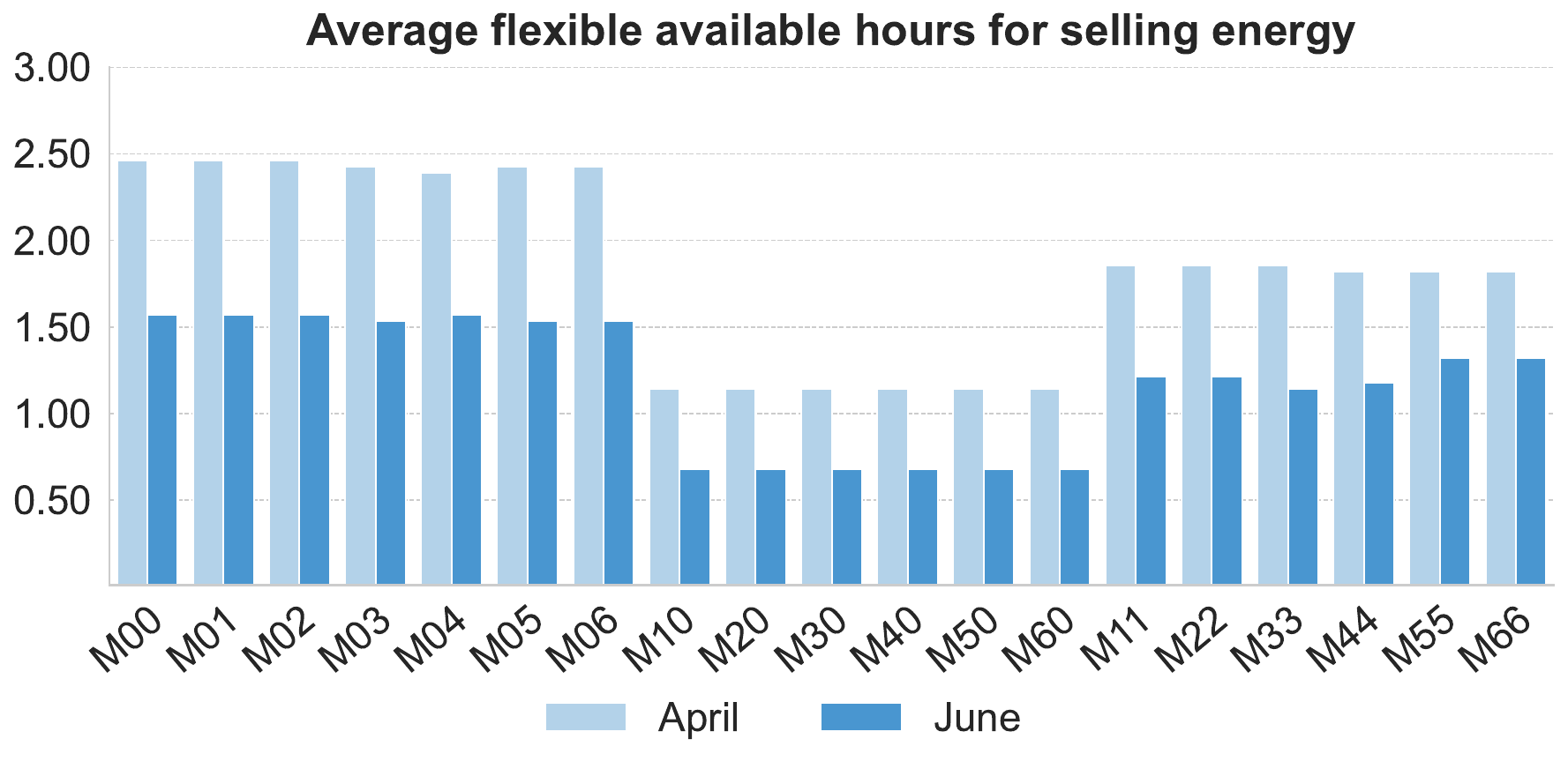}
\includegraphics[width=.45\textwidth]{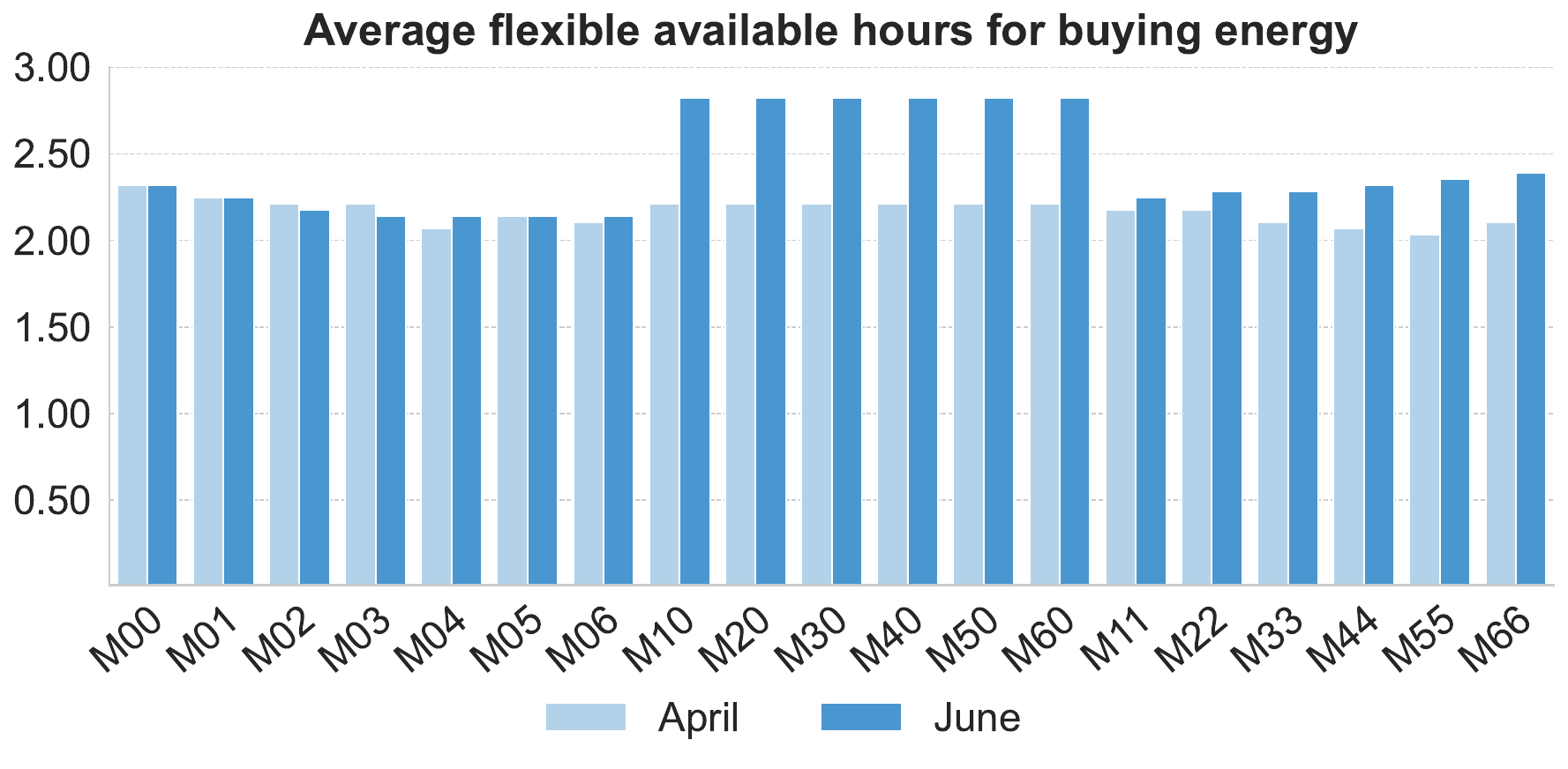}
\caption{Average daily hours with technically feasible flexibility actions. Battery storage increases flexibility windows across scenarios.}
\label{fig:flex_hrs}
\end{figure}

Flexibility depends on both technical feasibility and economic opportunity. Figure~\ref{fig:flex_hrs} presents the average number of hours per day in which a technically feasible flexibility action could be scheduled—i.e., hours where a load deviation (positive or negative) can be implemented without violating process or asset constraints.

\begin{itemize}
  \item \textbf{Buying flexibility is more prevalent} than selling, with technical feasibility observed in 2.0–2.7 hours/day versus 0.7–2.5 hours/day, respectively.
  \item This asymmetry stems from easier ramp-up operations (e.g., activating raw mills or charging batteries) and looser constraints on load increases compared to curtailments.
\end{itemize}

\textbf{Battery systems} notably expand the set of technically feasible hours for both regulation directions by enabling temporal decoupling. In contrast, \textbf{PV systems} have limited impact on short-term flexibility availability, as they mainly influence baseline cost optimization rather than moment-to-moment scheduling headroom.

\begin{figure}[h]
\centering
\includegraphics[width=.45\textwidth]{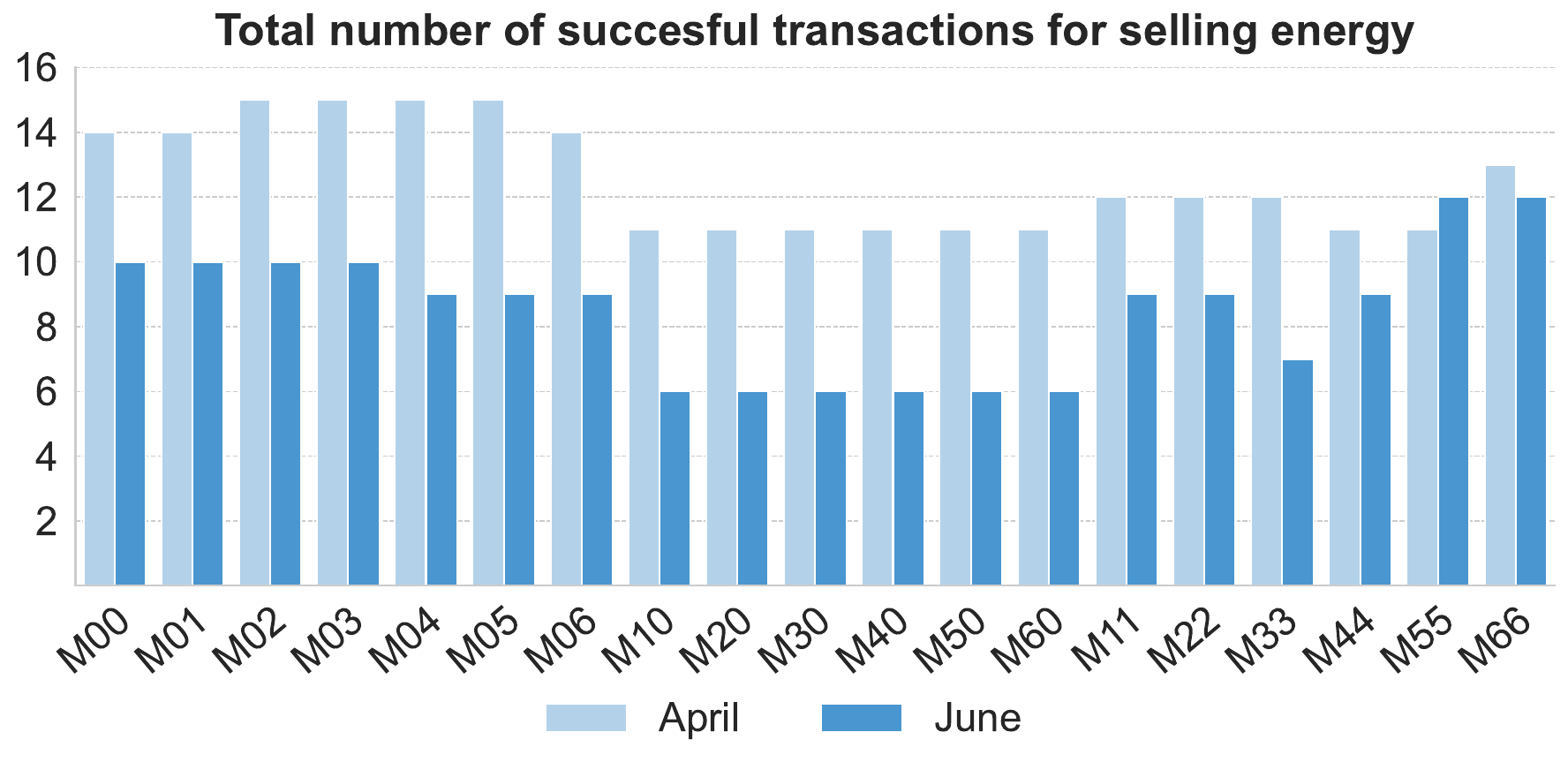}
\includegraphics[width=.45\textwidth]{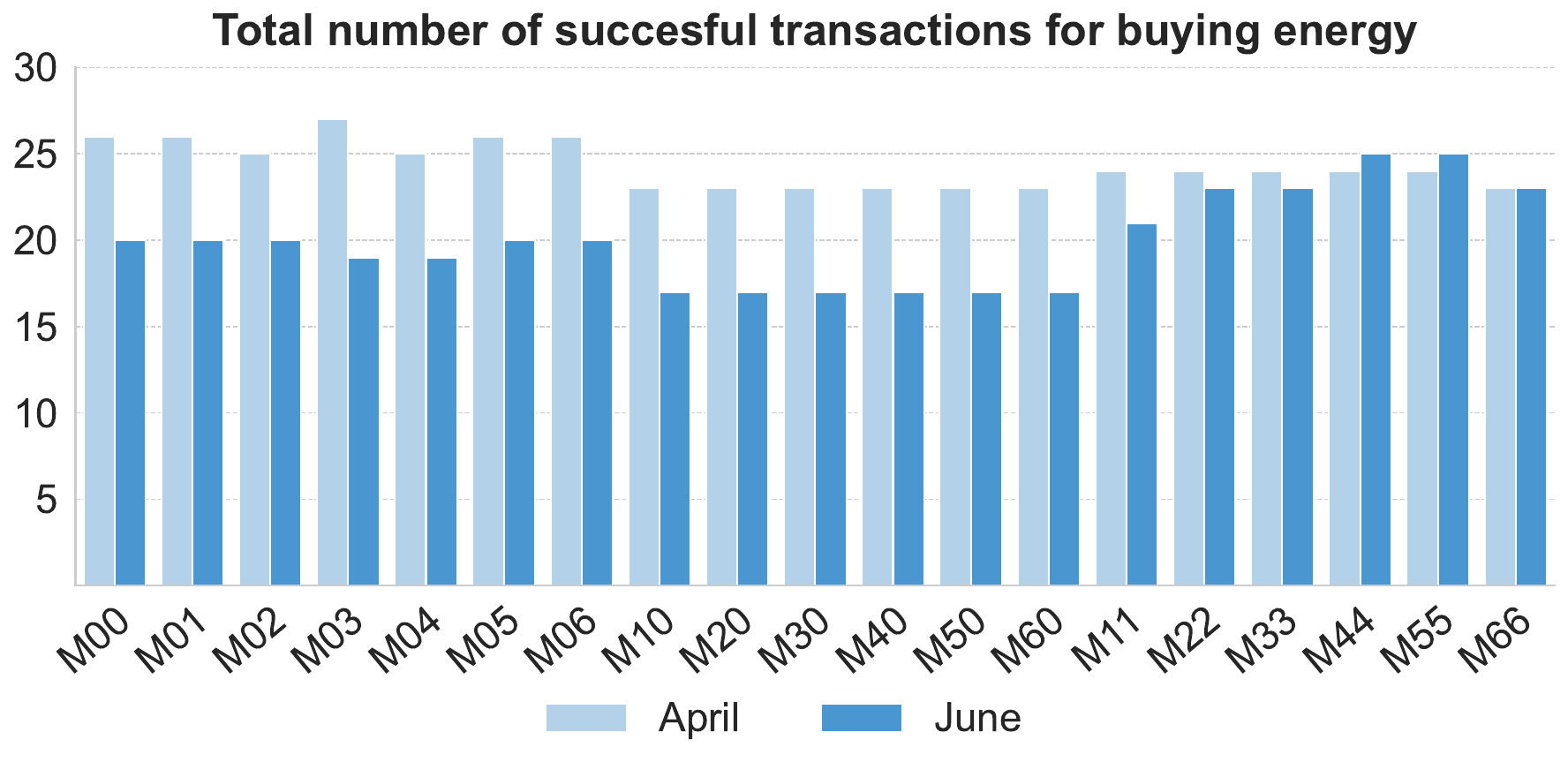}
\caption{Monthly accepted (i.e., technically and economically viable) flexibility transactions. More buying actions are accepted due to favorable market spreads and fewer operational constraints.}
\label{fig:success_trans}
\end{figure}

However, not all technically feasible actions are economically worthwhile. Figure~\ref{fig:success_trans} shows the number of \emph{successful transactions}—i.e., actions that are both feasible and profitable after comparing balancing market prices against the cost of deviating from the optimized baseline schedule.

\begin{itemize}
  \item \textbf{Buying:} 15–26 accepted transactions/month.
  \item \textbf{Selling:} 6–15 accepted transactions/month.
\end{itemize}

This difference highlights that market signals play a decisive role in monetizing flexibility. Increasing consumption during overgeneration periods (buying) tends to offer more consistent and higher spreads, especially when combined with the operational ease of ramping up equipment or charging batteries.

\subsection{Savings from Optimized Baseline Scheduling}

\begin{figure}[h]
\centering
\includegraphics[width=.45\textwidth]{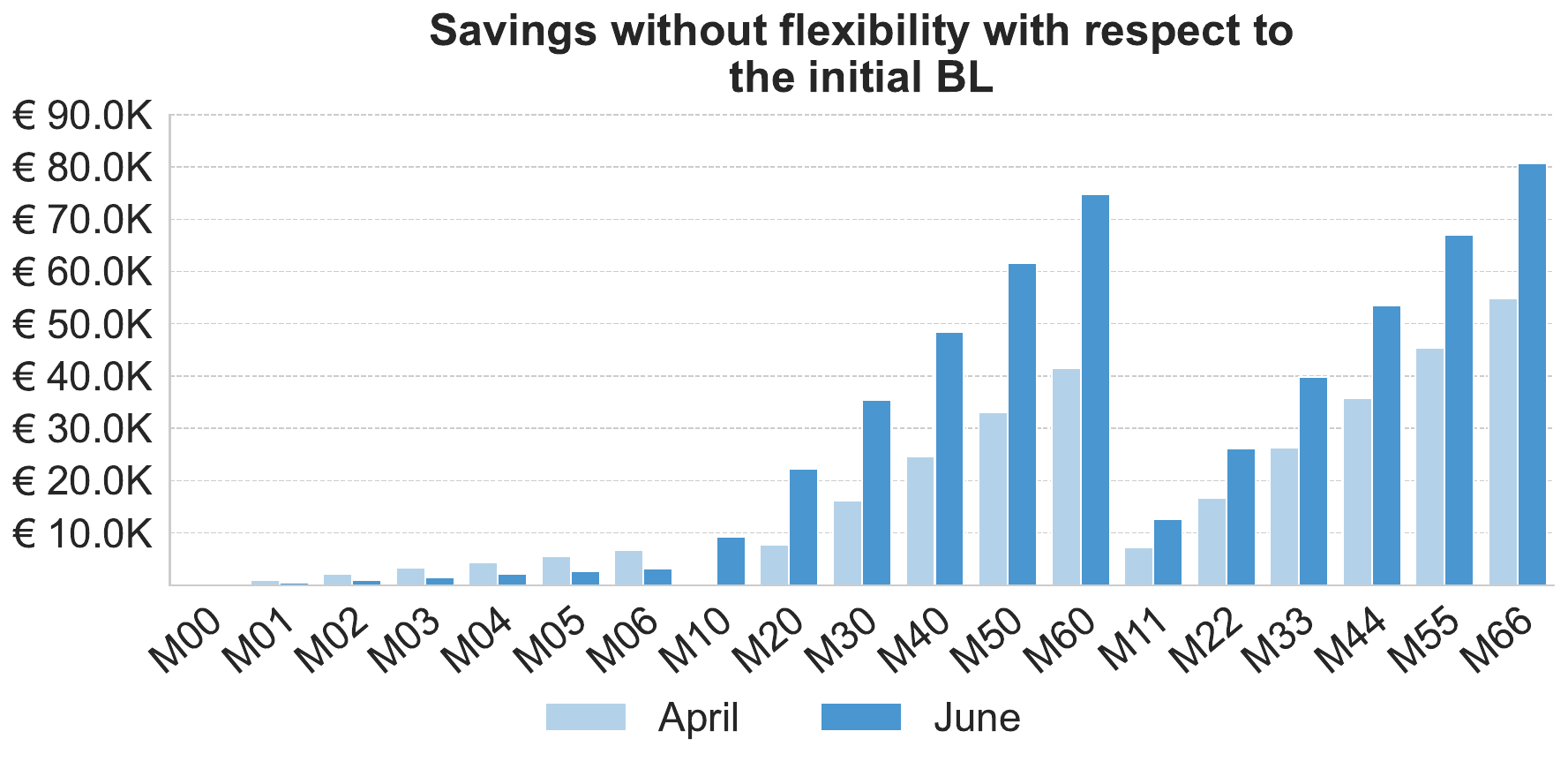}
\caption{Electricity cost savings from improved baseline scheduling, relative to the reference case (\texttt{M00}). PV dominates cost avoidance; BESS provides complementary returns.}
\label{fig:saving_noflex}
\end{figure}

Flexibility revenues are only part of the economic benefit. Energy assets also enable more efficient baseline scheduling. Figure~\ref{fig:saving_noflex} compares each scenario’s total electricity cost against the no-asset reference case.

\begin{itemize}
  \item \textbf{PV systems} offer significant cost reductions by directly displacing grid imports during sunny hours.
  \item \textbf{BESS units} allow intra-day arbitrage—charging during low-price periods and discharging during peaks—but offer modest savings on their own.
  \item \textbf{Combined PV+BESS scenarios} yield the greatest cost reductions, although diminishing returns appear at higher capacities.
\end{itemize}

\subsection{Investment Payback Period}

\begin{figure}[h]
\centering
\includegraphics[width=.45\textwidth]{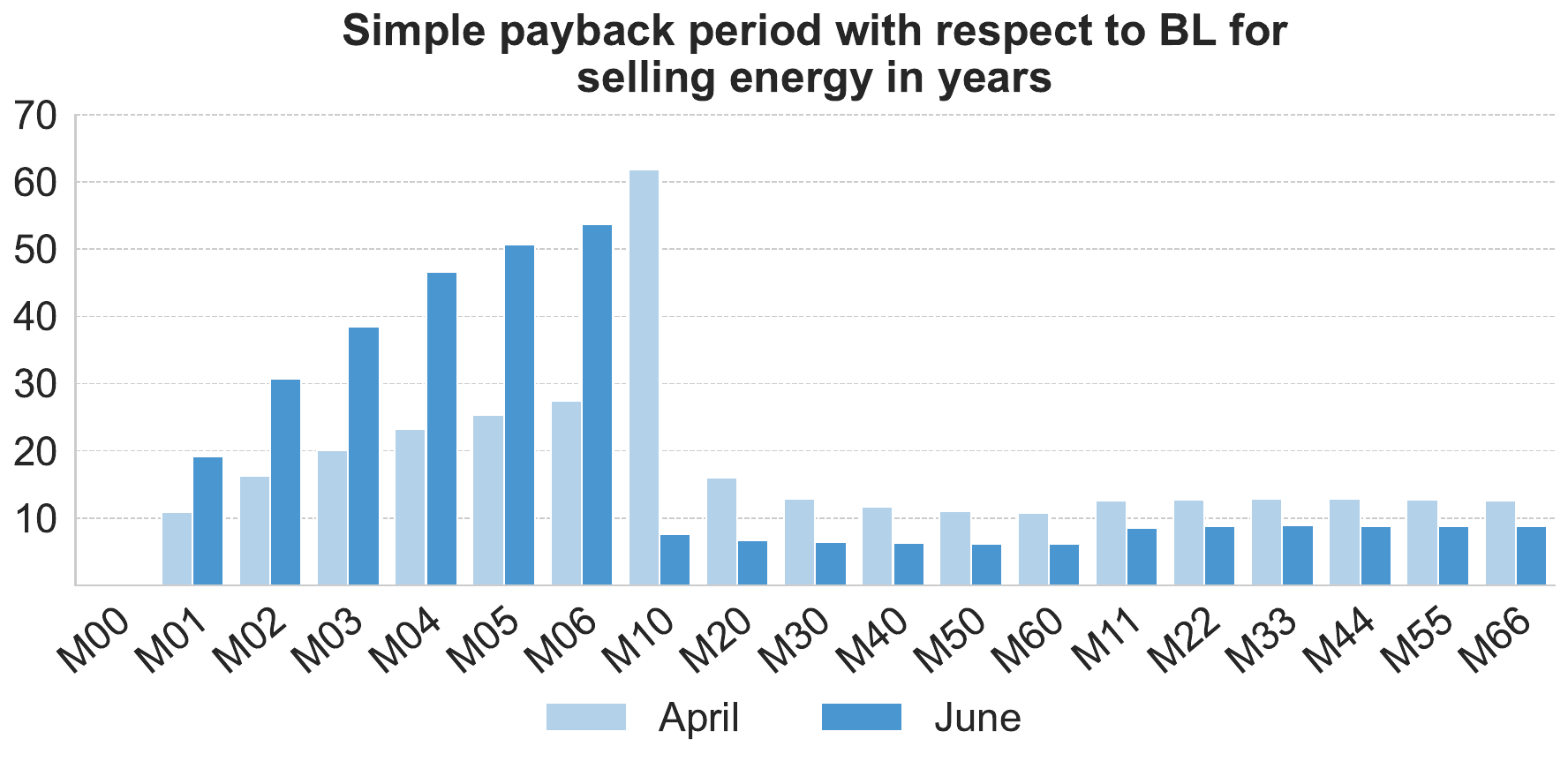}
\includegraphics[width=.45\textwidth]{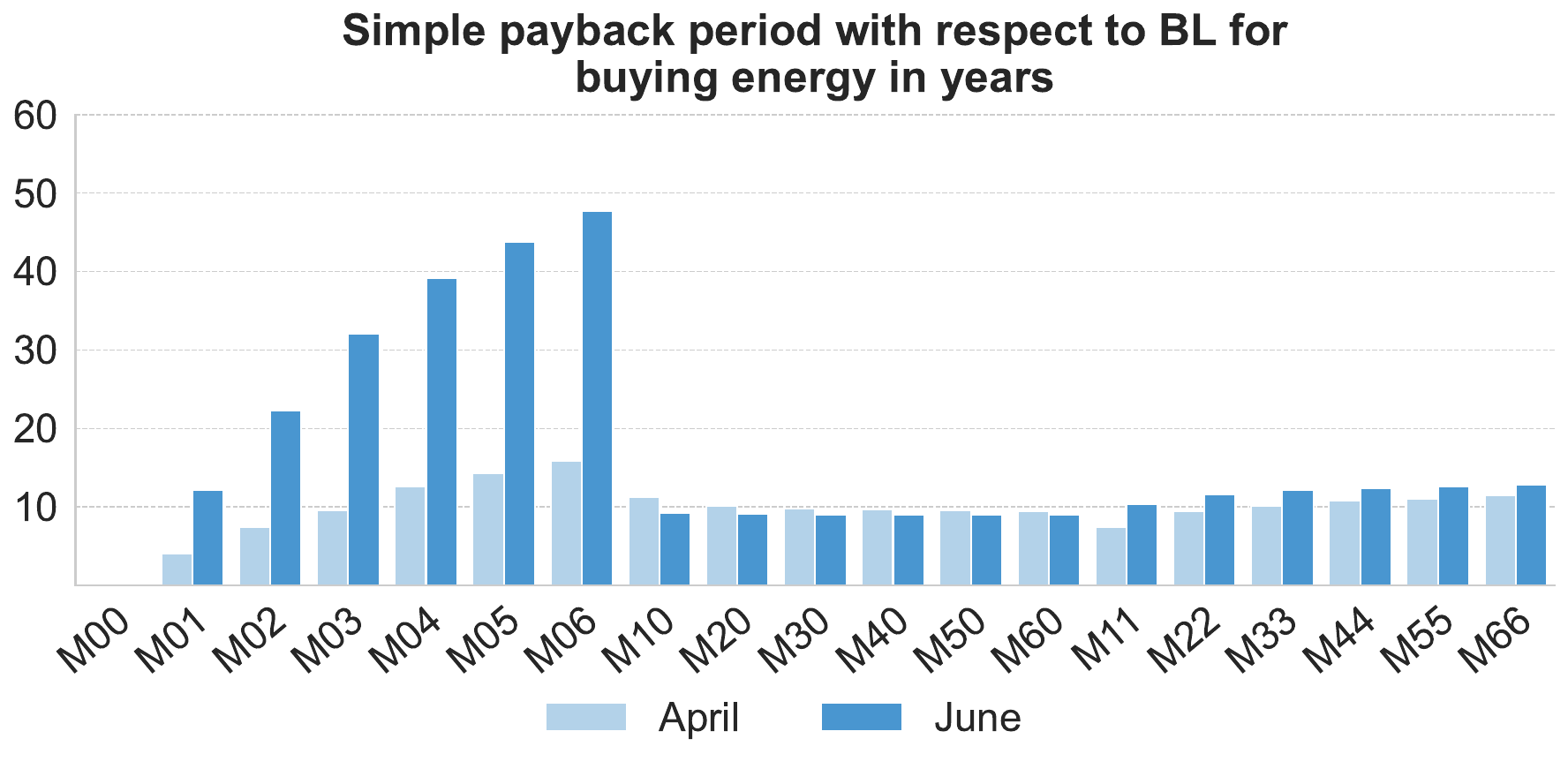}
\caption{Estimated investment payback periods. Best outcomes occur in PV+BESS configurations under favorable market conditions.}
\label{fig:payback_total}
\end{figure}

Figure~\ref{fig:payback_total} estimates payback periods using CAPEX assumptions from Lazard~\cite{LCOE_Lazard}:

\begin{itemize}
  \item PV: \SI{934500}{\euro\per\mega\watt}
  \item BESS: \SI{530885}{\euro\per\mega\watt\hour}
\end{itemize}

Annual savings are computed as 12 times the monthly total (baseline savings + flexibility revenues). This yields:

\begin{itemize}
  \item \textbf{PV-only systems}: 7–12 year paybacks, depending on market conditions and asset size.
  \item \textbf{PV+BESS combinations}: 5–7 years in optimal configurations, especially during high-spread periods (e.g., April).
  
\end{itemize}

\medskip
In conclusion, the results show that industrial electricity flexibility can be both technically viable and economically attractive—especially when paired with renewables and storage. Flexibility alone does not justify investment, but integrated strategies offer robust benefits, short paybacks, and alignment with decarbonization goals.

\section{Conclusions and Future Work}
\label{sec:conclusion}

This study presents a practical and computationally efficient methodology to identify and monetize short-term flexibility in industrial electricity consumption—bridging the gap between production planning and participation in balancing markets. A two-stage optimization framework was formulated and tested on a real-world Portland cement plant in Spain. The approach integrates cost-optimal production scheduling with flexibility assessment for participation in Spain’s tertiary regulation (manual Frequency Restoration Reserve, mFRR) market. On-site photovoltaic (PV) systems and battery energy storage systems (BESS) were modeled as controllable assets that contribute to both cost reduction and grid-responsive operation.

\subsection*{Key Findings}

The main findings can be summarized as follows:

\begin{itemize}
  \item \textbf{Baseline cost savings.} The Stage~1 model consistently achieved reduction in electricity costs by scheduling mill operation during low-price hours, maximizing PV self-consumption, and exploiting silo storage for demand buffering.

  \item \textbf{Viable market-based flexibility.} Stage~2 identified profitable flexibility opportunities in \textbf{18–30\%} of simulated hours. Monthly revenues ranged from \EUR{500} to \EUR{800}, depending on market conditions and asset configuration.

  \item \textbf{Added value of battery storage.} BESS improved flexibility outcomes by enabling larger deviations, increasing the number of accepted bids, and lowering the marginal cost of schedule adjustments.

  \item \textbf{Favorable investment metrics.} A combined configuration of \SI{1}{MWp} PV and \SI{1}{MWh} BESS achieved a \textbf{simple payback of 6–8 years} under April 2023 price conditions, validating the economic viability of the proposed strategy.

  \item \textbf{Synergistic asset contributions.} PV systems reliably reduced baseline costs, while BESS provided temporal flexibility. Together, they yielded superior outcomes compared to either asset alone.

  \item \textbf{Asymmetric flexibility.} Technically feasible buying actions (e.g., load increases) were more frequent than selling, and also more likely to be economically successful. This reflects both market asymmetries and the operational ease of up-ramping in process systems.

  \item \textbf{Primary value from scheduling.} The bulk of the economic gains originated from improved baseline scheduling. Flexibility revenues provided an important but secondary source of value.

  \item \textbf{Sensitivity to market context.} Profitability varied significantly between April and June, illustrating the importance of adapting strategies to market-specific conditions and seasonal price patterns.

  \item \textbf{Methodological robustness.} Consistent performance across contrasting market months (April and June) illustrates the framework's resilience and adaptability to different regulatory and pricing environments.
\end{itemize}

These results go beyond the specific case of cement manufacturing. They illustrate how industrial facilities can evolve from passive energy consumers to active players in power system flexibility. By combining intelligent scheduling, renewable integration, and market participation, the proposed methodology lays the groundwork for scalable, sector-wide strategies that enhance both economic performance and grid resilience—contributing directly to broader decarbonization goals.

\subsection*{Limitations}

While the framework demonstrated strong performance, several simplifying assumptions should be acknowledged:

\begin{itemize}
  \item \textbf{Deterministic inputs.} All simulations assumed perfect foresight for electricity prices, solar generation, and process demand. Forecast uncertainty was not modeled.

  \item \textbf{No asset degradation.} The economic evaluation did not account for equipment degradation—particularly battery cycling costs or wear-and-tear on raw mills.

  \item \textbf{Price-taker assumption.} The plant was assumed to operate without influencing market prices, which may not hold in low-liquidity periods or smaller balancing markets.
\end{itemize}

These assumptions introduce potential sources of optimistic bias, particularly in the estimation of flexibility revenues and system performance. Future research should incorporate stochastic elements and degradation-aware modeling to increase realism and robustness.

\subsection*{Future Research Directions}

Several avenues for further research and model enhancement are suggested:

\begin{itemize}
  \item \textbf{Stochastic and robust optimization,} to address uncertainty in price forecasts, PV output, and demand. This would improve the realism and reliability of scheduling and bidding strategies.

  \item \textbf{Multi-period and multi-market participation,} including secondary reserves and continuous intraday markets, to expand flexibility potential and assess cumulative value.

  \item \textbf{Multi-objective formulations,} to jointly optimize for economic cost, emissions reduction, and renewable integration using Pareto efficiency or weighted goals.

  \item \textbf{Degradation-aware modeling,} incorporating lifecycle effects and operational wear into scheduling decisions—especially for BESS and high-load machinery.

  \item \textbf{Broader sectoral applications,} testing the framework across other energy-intensive industries such as steel, chemicals, or pulp and paper, to generalize findings and support wider adoption.
\end{itemize}

\noindent
In conclusion, this study demonstrates that energy-intensive industrial facilities can actively contribute to grid balancing and decarbonization, while improving their economic performance. By integrating intelligent scheduling with on-site energy assets, industries can unlock new revenue streams and support a more flexible and sustainable power system.


\nomenclature[V]{$P_{b,t}$}{Power purchased from the grid at time $t$ [\unit{\mega\watt}].}
\nomenclature[V]{$P_{s,t}$}{Power exported to the grid at time $t$ [\unit{\mega\watt}].}
\nomenclature[V]{$P_{\mathrm{PV},t}$}{Power generated by the PV system at time $t$ [\unit{\mega\watt}].}
\nomenclature[V]{$P_{C,t}$, $P_{D,t}$}{Battery charge/discharge power at time $t$ [\unit{\mega\watt}].}
\nomenclature[V]{$Y_{k,t}$}{Binary on/off status of mill $k$ at time $t$.}
\nomenclature[V]{$I_{k,t}$}{Inventory in silo of mill $k$ at time $t$ [\unit{\tonne}].}
\nomenclature[V]{$\mathrm{SOC}_t$}{Battery state of charge at time $t$ [\unit{\mega\watt\hour}].}
\nomenclature[V]{$\Phi$}{Total electricity procurement cost over the planning horizon [\unit{\euro}].}
\nomenclature[V]{$\Phi^*$}{Electricity cost of the baseline schedule [\unit{\euro}].}
\nomenclature[V]{$\Phi^\dagger$}{Electricity cost of the flexibility-adjusted schedule [\unit{\euro}].}
\nomenclature[V]{$\Delta \Phi$}{Incremental cost from deviating the baseline schedule [\unit{\euro}].}
\nomenclature[V]{$\Delta P_\tau$}{Enforced power deviation in the balancing market [\unit{\mega\watt}].}
\nomenclature[V]{$D_t$}{Total process demand for raw meal at time $t$ [\unit{\tonne\per\hour}].}
\nomenclature[V]{$D_{k,t}$}{Portion of $D_t$ withdrawn from silo $k$ [\unit{\tonne\per\hour}].}

\nomenclature[C]{$\pi_t$}{Day-ahead electricity price at time $t$ [\unit{\euro\per\mega\watt\hour}].}
\nomenclature[C]{$\lambda_\tau^{\text{mFRR}\,+}$, $\lambda_\tau^{\text{mFRR}\,-}$}{Upward/downward mFRR price at time $\tau$ [\unit{\euro\per\mega\watt\hour}].}
\nomenclature[C]{$\lambda_\tau^{\text{spread}}$}{Required spread for flexibility profitability at time $\tau$ [\unit{\euro\per\mega\watt\hour}].}
\nomenclature[C]{$P_k$}{Rated power consumption of mill $k$ [\unit{\mega\watt}].}
\nomenclature[C]{$\Pi_k$}{Material throughput of mill $k$ [\unit{\tonne\per\hour}].}
\nomenclature[C]{$I_{k,\min}$, $I_{k,\max}$}{Minimum and maximum inventory in silo $k$ [\unit{\tonne}].}
\nomenclature[C]{$C_{\max}$}{Battery energy capacity [\unit{\mega\watt\hour}].}
\nomenclature[C]{$P_C^{\max}$, $P_D^{\max}$}{Maximum battery charge/discharge power [\unit{\mega\watt}].}
\nomenclature[C]{$P_b^{\max}$}{Maximum allowed grid import power [\unit{\mega\watt}].}
\nomenclature[C]{$\mathrm{DoD}$}{Battery depth of discharge [dimensionless].}
\nomenclature[C]{$\mathrm{SOC}_0$}{Initial battery state of charge [\unit{\mega\watt\hour}].}
\nomenclature[C]{$M_k^{\mathrm{ON}}$, $M_k^{\mathrm{OFF}}$}{Minimum on/off durations for mill $k$ [\unit{\hour}].}
\nomenclature[C]{$E_b^*$}{Total baseline grid energy imported [\unit{\mega\watt\hour}].}
\nomenclature[C]{$\varepsilon^-$, $\varepsilon^+$}{Lower and upper bounds for total energy deviation [dimensionless].}

\nomenclature[P]{$t$}{Time period index.}
\nomenclature[P]{$\tau$}{Activation time for flexibility.}
\nomenclature[P]{$\mathcal{T}$}{Set of time periods in the planning horizon.}
\nomenclature[P]{$\mathcal{K}$}{Set of raw milling units.}
\nomenclature[P]{$\mathcal{T}_{\mathrm{act}}$}{Set of candidate activation hours.}

\printnomenclature


\section*{Author contributions}
\textbf{Sebastián Rojas-Innocenti:} Conceptualization, Data Curation, Formal analysis, Investigation, Methodology, Resources, Software, Visualization, Validation, Writing - Original Draft, Writing - Review \& Editing.
\textbf{Enrique Baeyens:} Conceptualization, Formal analysis, Investigation, Methodology, Resources, Supervision, Validation, Visualization, Writing - Review \& Editing.
\textbf{Alejandro Martín-Crespo:} Conceptualization, Funding acquisition, Methodology, Project administration, Resources, Supervision, Validation, Writing - Review \& Editing, Visualization.
\textbf{Sergio Saludes-Rodil:} Conceptualization, Funding acquisition, Methodology, Project administration, Validation, Resources, Writing - Review \& Editing, Supervision. 
\textbf{Fernando Frechoso-Escudero:} Conceptualization, Methodology, Resources, Supervision, Writing - Review \& Editing, Visualization.

All authors have read and approved the final version of the manuscript.

\section*{Acknowledgments}
We appreciate the assistance of Fortia Energía for providing the related information on the Industrial Case Study.

\section*{Financial disclosure}
This research was founded by the MIG-20211033 grant from the Center for Industrial Technological Development (CDTI) of the Ministry of Science, Innovation, and Universities of the Spanish Government.
It has also been supported by Fundación CARTIF, a private and non-profit multidisciplinary Research Institution.

\section*{Conflict of interest}
The authors and Fundación CARTIF have no relevant financial or non-financial interests to disclose.

\section*{Data Availability}
The operational and market data used in this study were provided by Fortia Energía under a non-disclosure agreement and are not publicly available. Interested researchers may request access for academic or collaborative purposes, subject to Fortia's approval and confidentiality constraints.

\bibliography{Bibliography}

\end{document}